\documentclass[twocolumn]{aastex631}
\usepackage{amsmath, float, textgreek, enumitem, booktabs, needspace,mathalpha,rotating, color}

\received{June 20, 2025}
\revised{September 4, 2025}
\accepted{September 22, 2025}

\submitjournal{ApJS}

\shorttitle{Metallicities from TRES and \texttt{uberMS}}
\shortauthors{Pass et al.}

\begin{document}
\widowpenalty=0
\clubpenalty=0
\title{Metallicities from High-Resolution TRES Spectra with \texttt{uberMS}:\\ Performance Benchmarks and Literature Comparison}

\author[0000-0002-1533-9029]{Emily K. Pass}
\affiliation{Kavli Institute for Astrophysics and Space Research, Massachusetts Institute of Technology, Cambridge, MA 02139, USA}

\author[0000-0002-1617-8917]{Phillip A. Cargile}
\affiliation{Center for Astrophysics $\vert$ Harvard \& Smithsonian, 60 Garden Street, Cambridge, MA 02138, USA}

\author[0000-0003-0741-7661]{Victoria DiTomasso}
\affiliation{Center for Astrophysics $\vert$ Harvard \& Smithsonian, 60 Garden Street, Cambridge, MA 02138, USA}

\author[0000-0003-1445-9923]{Romy Rodríguez Martínez}
\affiliation{Center for Astrophysics $\vert$ Harvard \& Smithsonian, 60 Garden Street, Cambridge, MA 02138, USA}

\author[0000-0002-9003-484X]{David Charbonneau}
\affiliation{Center for Astrophysics $\vert$ Harvard \& Smithsonian, 60 Garden Street, Cambridge, MA 02138, USA}

\author[0000-0001-9911-7388]{David W. Latham}
\affiliation{Center for Astrophysics $\vert$ Harvard \& Smithsonian, 60 Garden Street, Cambridge, MA 02138, USA}

\author[0000-0001-7246-5438]{Andrew Vanderburg}
\affiliation{Kavli Institute for Astrophysics and Space Research, Massachusetts Institute of Technology, Cambridge, MA 02139, USA}

\author[0000-0001-6637-5401]{Allyson Bieryla}
\affiliation{Center for Astrophysics $\vert$ Harvard \& Smithsonian, 60 Garden Street, Cambridge, MA 02138, USA}

\author[0000-0002-8964-8377]{Samuel N. Quinn}
\affiliation{Center for Astrophysics $\vert$ Harvard \& Smithsonian, 60 Garden Street, Cambridge, MA 02138, USA}

\author[0000-0003-1605-5666]{Lars A. Buchhave}
\affiliation{DTU Space, National Space Institute, Technical University of Denmark, Elektrovej 328, DK-2800 Kgs.\ Lyngby, Denmark}

\begin{abstract}
\noindent As the field of exoplanetary astronomy has matured, demand has grown for precise stellar abundances to probe subtle correlations between stellar compositions and planetary demographics. However, drawing population-level conclusions from the disparate measurements in the literature is challenging, with various groups measuring metallicities using bespoke codes with differing line lists, radiative transfer models, and other assumptions. Homogeneous analyses are thus critical. Here we use the neural-net framework \texttt{uberMS} to measure iron abundances and alpha enrichments from high-resolution optical spectra observed by the Tillinghast Reflector Echelle Spectrograph (TRES), a key resource used for the follow-up of candidate exoplanet hosts. To contextualize these measurements and benchmark our method's performance, we compare to external constraints on metallicity using the Hyades cluster, wide binaries, and asteroids, to external constraints on $T_{\rm eff}$ and $\log g$ using stars with interferometric radii, and to the results of other abundance measurement methods using overlap samples with the APOGEE and SPOCS catalogs, as well as by applying the \texttt{SPC} method directly to TRES spectra. We find that TRES--\texttt{uberMS} provides reliable parameter estimates with errors of roughly 100~K in $T_{\rm eff}$, 0.09~dex in $\log g$, and 0.04~dex in [Fe/H] for many nearby dwarf stars, although [Fe/H] performance is poorer for mid-to-late K dwarfs, with the bias worsening with decreasing $T_{\rm eff}$. Performance is also worse for evolved stars. For [\textalpha/Fe], our error may be as good as 0.03~dex for dwarfs based on external benchmarks, despite sizable and variable systematic differences when comparing with specific alpha-element abundances from other catalogs.
\end{abstract}

\section{Introduction} \label{sec:intro}
Stellar metallicity is a key property in the era of exoplanet searches, both because precise stellar parameters are needed to obtain precise planetary parameters \citep[e.g.,][]{Muirhead2012, Tayar2022} and because stellar metallicities are correlated with the presence of planets \citep{Fischer2005, Dawson2013, Buchhave2018}. In addition, recent works are striving to unveil subtle and profound correlations between stellar composition and planetary demographics, such as comparing planetary and stellar iron-to-silicate mass fractions \citep{Plotnykov2020, Adibekyan2024}, investigating differences in planet populations between the galactic thin disk, thick disk, and halo \citep{Bashi2022}, and exploring links between metallicity and a diverse array of planetary properties, such as eccentricity \citep{An2023}, mutual inclination \citep{Hua2025}, system architecture \citep{Zhu2024}, and the obliquity of hot Jupiters \citep{Spalding2022}, among many other applications.

These science cases require measurement of accurate and precise stellar metallicities, an industry that began a century ago with Cecilia Payne (later Payne-Gaposchkin), who measured individual line strengths by hand \citep{Payne1925}. While modern abundance inference methods are considerably more advanced, they are not necessarily consistent with each other: contemporary methods must choose between numerous competing model atmospheres, line lists, and radiative transfer codes, as well as additional assumptions such as 3D behaviors, non-LTE effects, and the solar composition \cite[for review, see][]{Jofre2019}. Homogeneous stellar analyses are thus critical for exoplanetary demographics work to ensure results are not affected by systematic differences between inference methods.

With this in mind, the goal of this work is to homogeneously measure metallicity and alpha enrichment for a large sample of high-resolution spectra of planet-host stars that have been collected as part of exoplanet validation efforts. Specifically, this work outlines the method and the various benchmarking tests that we perform in order to evaluate the uncertainties in our abundance measurements. Subsequent papers from our team will apply the technique in a variety of science cases, including TESS planet demographics, hot Jupiters in the thick disk, and the planets of low-mass M dwarfs.

In Section~\ref{sec:uberms}, we outline \texttt{uberMS}, the neural-net framework that we use to measure stellar parameters. In Section~\ref{sec:tres}, we describe the TRES spectroscopy that we seek to fit; Section~\ref{sec:app} provides the details of that implementation. In Section~\ref{sec:results}, we describe the results of our various performance benchmarking tests: the Hyades open cluster in Section~\ref{sec:hyades}, the Gaia Benchmark Sample of stars with interferometric radii in Section~\ref{sec:gbs3}, the large-scale APOGEE survey in Section~\ref{sec:apogee}, the planet-search sample SPOCS in Section~\ref{sec:spocs}, asteroids in Section~\ref{sec:asteroids}, and wide stellar binaries in Section~\ref{sec:binaries}. We conclude with a summary in Section~\ref{sec:summary}.

\section{Methods}
\subsection{\texttt{uberMS}}
\label{sec:uberms}
We use the stellar inference code \texttt{uberMS},\footnote{\href{hhttps://github.com/pacargile/uberMS}{https://github.com/pacargile/uberMS}} designed to efficiently and accurately determine precise stellar parameters within a full Bayesian framework. A forthcoming paper (Cargile et al.\ in prep) will present a detailed description of this code. Here we briefly outline its principal methodology and underlying stellar models.

The \texttt{uberMS} code builds upon the legacy of \texttt{The Payne} \citep{Ting2019} and \texttt{MINESweeper} \citep{Cargile2020}, leveraging their neural-network emulation methods to facilitate rapid and reliable stellar parameter inference. The primary distinction between \texttt{uberMS} and its predecessors lies in the inference technique—while the underlying stellar models and emulation approaches remain largely consistent, \texttt{uberMS} introduces the use of Stochastic Variational Inference \citep[SVI; ][]{Phan2019,Bingham2019} as its core Bayesian inference engine.

\texttt{uberMS} integrates observational stellar photometry and spectroscopy with theoretical stellar atmosphere models, adopting a flexible hierarchical Bayesian structure capable of incorporating informative priors. Posterior distributions of fundamental stellar parameters—effective temperature ($T_{\mathrm{eff}}$), surface gravity ($\log g$), metallicity ([Fe/H]), alpha-element enhancement ([\ensuremath{\alpha}/Fe]), stellar radius ($\log R$), rotational and macroturbulence broadening ($v_*$), stellar mass, age, and distance—are inferred using SVI, a computationally efficient probabilistic inference technique.

SVI approximates complex posterior distributions by optimizing a simpler, parameterized variational distribution to closely match the true posterior. In \texttt{uberMS}, Normalizing Flows—a family of neural-network-based transformations—are employed to transform a Gaussian base distribution into a complex distribution which accurately approximates the true stellar parameter posterior shapes. Unlike classical Markov Chain Monte Carlo (MCMC) or nested sampling methods, which generate posterior samples through computationally demanding iterative sampling, SVI employs stochastic optimization to iteratively update the variational parameters by computing gradients of the evidence lower bound (ELBO) of the posterior. This optimization approach significantly reduces computational time over traditional sampling techniques.

Within \texttt{uberMS}, the likelihood of an observed stellar spectrum and/or photometry is computed using synthetic data predicted by \texttt{The Payne}, a neural-network-based spectral emulator trained on grids of theoretical synthetic spectra. This extensive spectral grid was computed using the Kurucz line list and associated codes, specifically the \texttt{ATLAS-12} model atmospheres \citep{Kurucz2005} and the \texttt{SYNTHE} radiative transfer code \citep{Kurucz1993}, spanning $T_{\mathrm{eff}}$ and $\log g$ across the full H-R diagram. The \citet{Grevesse1998} solar-abundance scale was adopted when computing this grid; this scale is in good agreement with the latest solar abundance measurements from \citet{Magg2022}. In addition, \texttt{uberMS} can optionally incorporate stellar isochrone priors based on stellar evolution models. It uses an approach analogous to that of \texttt{MINESweeper}, employing a neural-network emulator to rapidly interpolate the MIST stellar isochrone grid \citep{Choi2016}, itself based on the MESA stellar evolution code \citep{Paxton2011}. Throughout this work, we refer to modeling the data with just the stellar atmospheres as running \texttt{uberMS} in `TP mode'; when isochrones are also included, we are using `MS mode.'

\subsection{TRES}
\label{sec:tres}
The Tillinghast Reflector Echelle Spectrograph (TRES) is an $R$=44,000 echelle spectrograph located at the 1.5m Tillinghast Reflector at the Fred Lawrence Whipple Observatory on Mt.\ Hopkins, AZ, spanning a wavelength range of 390--910nm \citep{Szentgyorgyi2007}. It saw first light in 2007 and has since been in regular use for a wide variety of investigations, resulting in a rich archive of high-resolution optical observations: as of 2025, the archive contains nearly 100,000 spectra. TRES is a major contributor to the TESS Follow-up Observing Program (TFOP;\footnote{\href{https://tess.mit.edu/followup/}{https://tess.mit.edu/followup/}} \citealt{Collins2018}) and hence most nearby exoplanet hosts accessible from the northern hemisphere have spectra in the TRES archive.

Spectra in the TRES archive have been reduced using a standard pipeline \citep{Buchhave2010}, which includes flat-fielding, cosmic-ray rejection, echelle order extraction, and wavelength calibration with ThAr spectra. In this work, we also subject the spectra to additional preprocessing, with many of these routines adapted from the \texttt{tres-tools} package by Jonathan Irwin.\footnote{\href{https://github.com/mdwarfgeek/tres-tools}{https://github.com/mdwarfgeek/tres-tools}} Using TRES order 23 (514--523nm), we cross-correlate each spectrum with a coadded template to shift the spectrum to the stellar rest frame, then convert from air to vacuum wavelengths. We refine the blaze correction using the nearly nightly TRES standard star observations taken of Sirius and/or Vega, fitting the change in shape relative to a coadded reference spectrum from 2013 using a 10th order Chebyshev polynomial. As we find that the changes in the blaze occur on long timescales, we take the rolling average of the corrections from within $\pm$1 week of our night of interest to increase the stability of the correction. After applying these preprocessing steps, we coadd all observations of a given star using an error-weighted mean. In cases where we have more than three spectra, we also apply 5$\sigma$ outlier clipping to this coaddition to resolve any cosmic-ray issues that were missed by the standard pipeline.

\subsection{Application of \texttt{uberMS} to TRES data}
\label{sec:app}
In this study, we apply \texttt{uberMS} specifically to spectra from the TRES archive. We employ a model that uses \texttt{The Payne} trained on synthetic spectra covering the Mg b triplet region (515–530nm). This model was initially developed for the H3 survey \citep{Conroy2019} and includes empirical tuning of the atomic and molecular transition line lists based on the observed solar and Arcturus spectra. This spectral region is particularly effective for robust stellar parameter determination due to the presence of pressure and alpha-abundance sensitive Mg lines complemented by numerous Fe I lines, which collectively provide strong constraints.

We consider TRES orders 23 and 24, as these orders overlap with the wavelength range for which the model has been trained. We crop the orders to this wavelength range, normalize to the median flux level in each order, and include the orders individually in the \texttt{uberMS} fits, allowing the various spectroscopic parameters to differ from order to order (namely, the spectral resolution, an RV offset, and the two parameters of a linear function that refine the continuum shape). We consider using a higher-order continuum polynomial, but we find that the higher-order terms can create degeneracies with stellar parameters, and the linear function yields reasonable agreement with the models after applying the Sirius/Vega blaze correction refinement described in the previous section. For the RV offset, note that our TRES preprocessing includes subtraction of the star's bulk RV; we therefore expect this offset to be close to zero.

For photometry, we consider magnitudes from 2MASS \citep[$J$, $H$, and $K_{\rm s}$; ][]{Skrutskie2006} and Gaia DR3 \citep[$G$, $G_{\rm BP}$, and $G_{\rm RP}$;][]{Gaia2023}, as this information is generally available for all stars we use in the various benchmarking tests in this work. \texttt{uberMS} has the flexibility to include any and all photometric data available for a source; however, to ensure consistency across our various tests, we opt to exclude other bands for which only some of our stars have extant photometry.

Table~\ref{tab:priors} gives the parameters/priors we use in our \texttt{uberMS} fits. We initialize from the same point for each star: for TP mode, this is $T_{\rm eff}=5000$~K, $\log g=4.5$~dex, [Fe/H]$=0.0$~dex, [\textalpha/Fe]$=0.0$~dex, $\log R=0.0$~dex, and $v_*=2$~kms$^{-1}$, while for MS mode the fit is performed in terms of the equivalent evolutionary phase (EEP; see \citealt{Dotter2016} for details), the initial [Fe/H] and [\textalpha/Fe], and the initial stellar mass instead of the present-day $T_{\rm eff}$, $\log g$, and surface abundances. Our prior on EEP prevents the fit from yielding evolutionary phases earlier than the zero-age main sequence or later than the tip of the red giant branch, with a heavier weighting towards the main sequence; we also add a prior on the latent age variable to penalize solutions that result in ages older than the age of the universe.

\begin{figure*}\centering\includegraphics[width=0.95\textwidth]{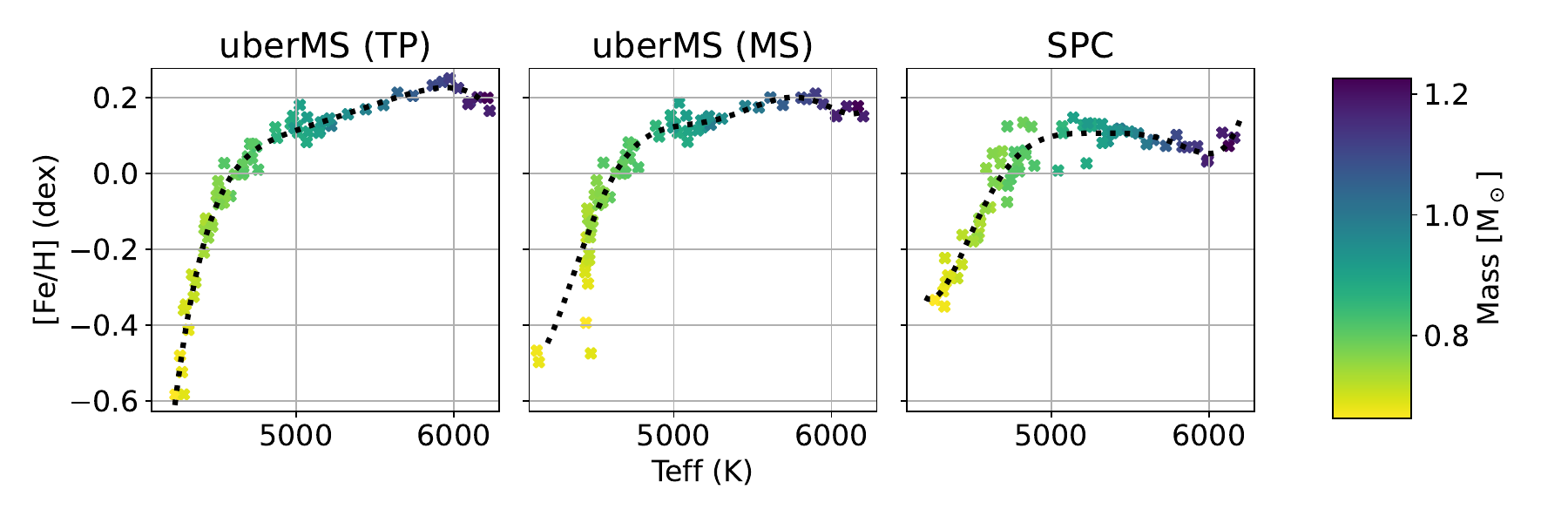}
    \caption{The metallicities and effective temperatures estimated from TRES spectra of 67 Hyades dwarf stars. From left to right, the panels show the results from \texttt{uberMS} in TP mode (no isochrones), \texttt{uberMS} in MS mode (including isochrones), and \texttt{SPC}. Points are color-coded based by stellar mass, as calculated from a Hyades-specific $K_{\rm s}$-band mass-luminosity relation. Black dotted lines represent sixth-order Legendre polynomial fits to the data, with coefficients given in Table~\ref{tab:leg}.}
    \label{fig:hyades}
\end{figure*}

Our choices of priors are determined through an iterative process, motivated by the performance found for the various benchmarking tests described in Section~\ref{sec:results} of this manuscript. One of these choices is to fix the spectroscopic jitter at 0.015 (in units of normalized flux). This value reflects an inferred noise floor in our spectroscopic data that results from systematic uncertainties in the stellar models and/or unaccounted for uncertainties in the TRES data; when specjitter is allowed to float, we find that multiple observations of the same target (and in particular, the asteroid sample that we will discuss in Section~\ref{sec:asteroids}) yield inconsistent inferred parameters, and by tuning this specjitter parameter we are able to bring the measurements into good agreement. Another variable that we constrain is the microturbulence, as we find that our data do not strongly inform this parameter. We therefore allow microturbulence to float, but place a deterministic prior constraining it to fall on the empirical relation from \citet{Bruntt2012} based on $T_{\rm eff}$ and $\log g$. Lastly, we place an informative prior on the extinction, weighting it towards low values; we make this choice because our observations do not strongly inform this parameter, but little extinction is expected given that we are considering nearby stars. The specific upper limit of $A_{\rm v}$=$0.1$ mag that we adopt is motivated by \citet{Soubiran2024}, who found that 90\% of stars in the Gaia Benchmark Sample (see Section~\ref{sec:gbs3}) have $A_{\rm v} < 0.05$ mag, with only five giants among their 201-star sample having $A_{\rm v} \geq 0.1$ mag; their findings are based on the 3D extinction maps of \citet{Vergely2022}.

\section{Literature Comparison}
\label{sec:results}
\subsection{SPC}
\label{sec:spc}
The Stellar Parameter Classification (\texttt{SPC}) tool \citep{Buchhave2012} is another spectral analysis code that has been commonly applied to TRES data. In particular, it has been used to systematically analyze reconnaissance spectra as part of TFOP, which vets stars for candidate exoplanets \citep{Bieryla2021}. To explore any differences in output parameters between this method and \texttt{uberMS}, we perform a complementary \texttt{SPC} analysis for each of the samples we investigate in this work. A key difference of \texttt{SPC} is that it fits for a bulk metallicity, [M/H], by assuming that individual metals appear in the same relative abundance pattern as in the Sun. The metallicity information that can be extracted by this method is therefore more limited compared to \texttt{uberMS}, which measures both [Fe/H] and [\textalpha/Fe].

Similar to \texttt{uberMS}, \texttt{SPC} cross-correlates the observed spectrum against a grid of \citet{Kurucz1992} models in the wavelength range 505--536nm. \texttt{SPC} uses Yonsei-Yale isochrones \citep{Yi2001} to set priors on $\log g$, a treatment that represents a middle ground between \texttt{uberMS}'s TP mode, which uses no isochrone information, and its MS mode, which directly fits isochrones to the observations. Differences between \texttt{SPC} and \texttt{uberMS} outputs are highlighted in the subsequent sections, with supplementary plots and discussion provided in Appendix~\ref{sec:app2}.

\begin{table*}[]
\noindent\caption{Coefficients of sixth-order Legendre polynomial fits to the Hyades metallicity estimates of Figure~\ref{fig:hyades}}
\label{tab:leg}
\begin{tabular}{@{}lrrrrrrr@{}}
\toprule
 & $c_0$ & $c_1$ & $c_2$ & $c_3$ & $c_4$ & $c_5$ & $c_6$ \\ \midrule
\texttt{uberMS} (TP) & 0.0736 & 0.2761 & $-$0.2129 & 0.1402 & $-$0.1161 & 0.0374 & $-$0.0160 \\
\texttt{uberMS} (MS) & 0.0695 & 0.2335 & $-$0.1983 & 0.0878 & $-$0.0523 & $-$0.0142 & 0.0408 \\
\texttt{SPC} & 0.0173 & 0.1673 & $-$0.1863 & 0.0859 & 0.0301 & $-$0.0165 & 0.0485 \\ \bottomrule
\end{tabular}
\tablecomments{\footnotesize These fits take the form $p(x) = c_0 * L_0(x) + c_1 * L_1(x) + \ldots + c_6 * L_6(x)$, as implemented in \texttt{numpy.polynomial.legendre.legval}. To improve the stability of the coefficients, $x$ is related to $T_{\rm eff}$ by the transformation $x = \frac{T_{\rm eff} - 5200}{1000}$; i.e., $x$ spans the interval [$-1$,+1].}
\end{table*}

\subsection{Hyades}
\label{sec:hyades}
Open clusters are useful testing grounds for metallicity estimation methods since their members share a common birth environment, age, and metallicity. Here we consider the Hyades cluster, with literature age estimates in the range 600--800Myr depending on the source \citep{Douglas2019}. The Hyades is well represented in the TRES archive as a result of the CfA Hyades binary survey, a long-term monitoring campaign to search for binary companions to Hyades stars that first began with the CfA digital speedometer and was continued with TRES \citep[e.g.,][]{Stefanik1992, Quinn2014, Torres2019}. Specifically, we identify 67 Hyades dwarf stars in the TRES archive that are not double-lined binaries, and analyze these stars using the methods described in Section~\ref{sec:app}. Our resulting [Fe/H] measurements as a function of effective temperature are given in Figure~\ref{fig:hyades}, alongside analogous results from \texttt{SPC}. The plotted values are the medians of the \texttt{uberMS} posteriors. While uncertainties can also be inferred from these posteriors, we find that they are severely underestimated as they do not reflect systematic errors in the theoretical models. We therefore do not include any error bars in the plot; of course, the goal of this work is to ultimately determine the appropriate error bars to use for each parameter.

The color bar in Figure~\ref{fig:hyades} gives the stellar mass, as inferred from a $K_{\rm s}$-band mass-luminosity relation for a 750Myr, [Fe/H]=0.18dex MIST model \citep{Choi2016}. While the true age and metallicity of the Hyades may deviate somewhat from these values, \citet{Torres2024b} showed that this model provides good agreement with the empirical $K_{\rm s}$-band mass-lumonisity relation that they determine using dynamical masses from 19 Hyades stars in astrometric binaries. Our calculation uses 2MASS $K_{\rm s}$-band magnitudes \citep{Skrutskie2006} and Gaia DR3 parallaxes \citep{Gaia2023}. This stellar mass provides an independent variable that we can use to select and compare subsamples of stars across different fitting methods; $T_{\rm eff}$ is not ideal for this purpose given that it is also an output of the fit and therefore varies between the subplots.

\begin{figure*}[p]
\centering
    \includegraphics[width=0.8\textwidth]{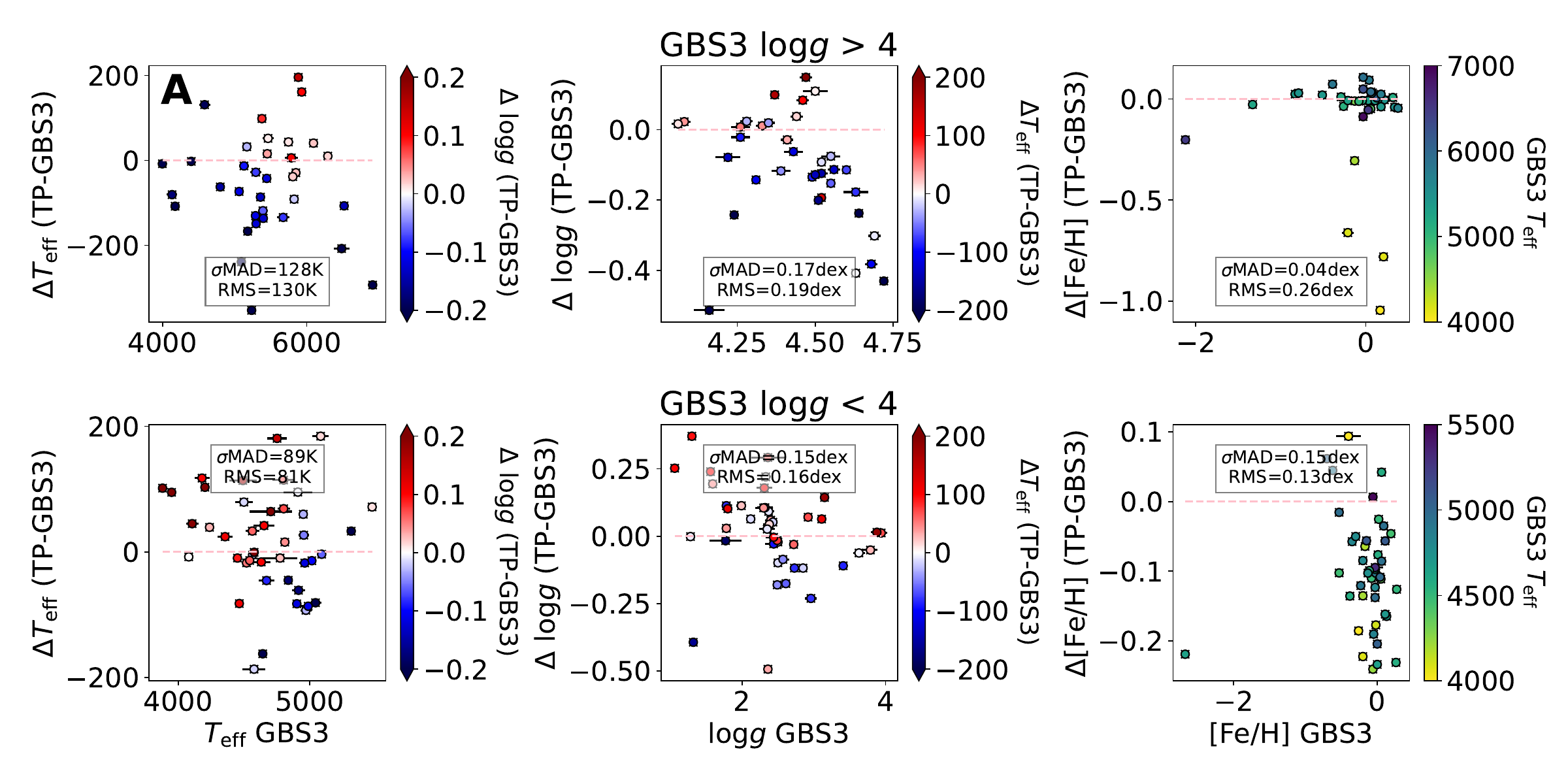}
    \includegraphics[width=0.8\textwidth]{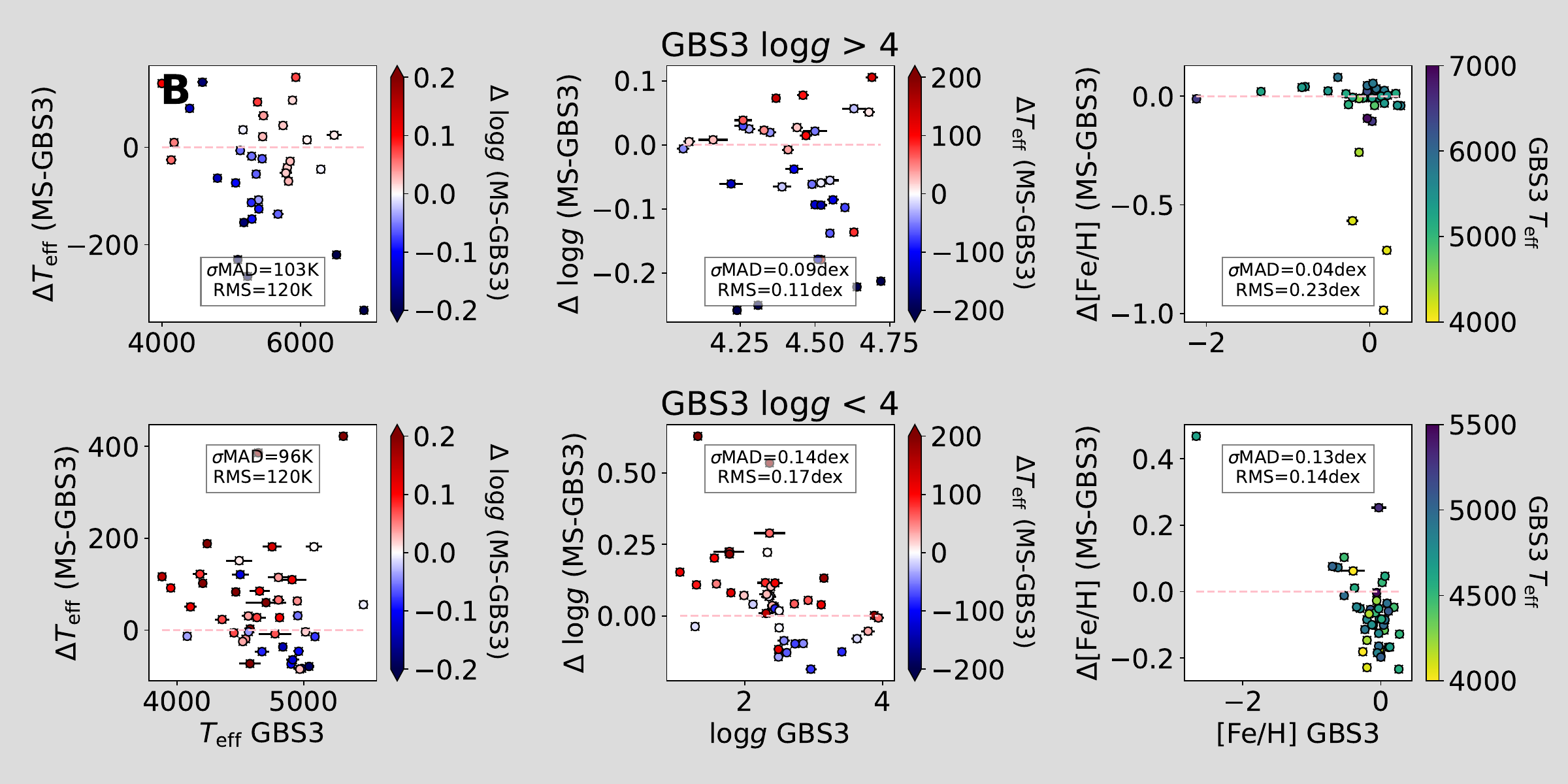}
    \includegraphics[width=0.8\textwidth]{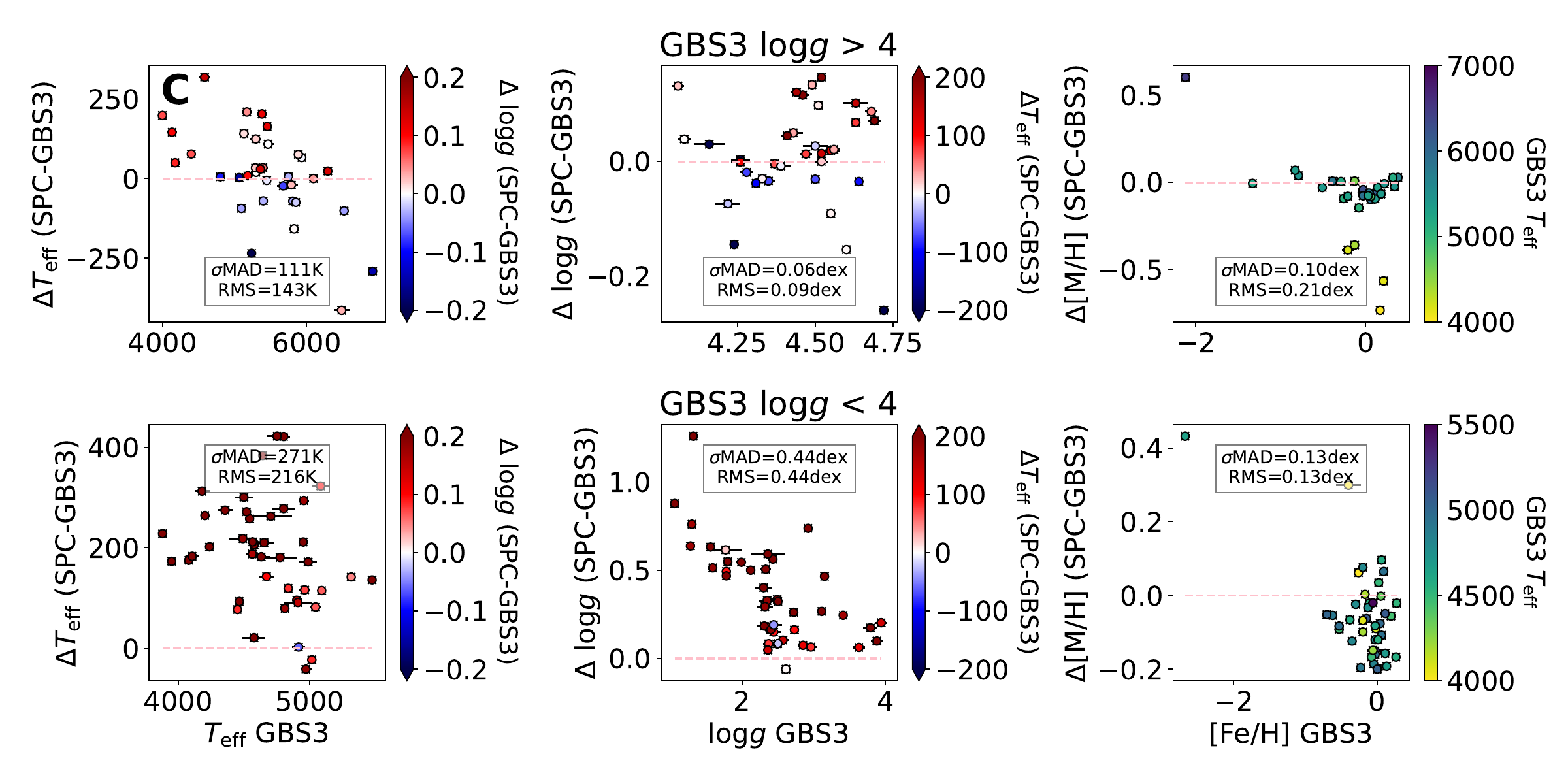}
    \caption{Comparison between GBS3 stellar parameters and properties estimated from TRES spectra using \texttt{uberMS} in TP mode (A), \texttt{uberMS} in MS mode (B), and \texttt{SPC} (C). In each set of subplots, the upper row shows dwarf stars (here approximated by log$g>4$ dex), while the lower row shows evolved stars (log$g<4$ dex). Note that error bars are only plotted for GBS3 values, not for our measurements. The dashed lines indicate unity.}
    \label{fig:gbs3}
\end{figure*}

The performance of all three methods is poor at cool temperatures: we find lower and lower metallicities for cooler and cooler stars, despite our expectation that the metallicity for all stars in the cluster should be the same. This phenomenon stems from the limitations of theoretical models, an issue that has long been appreciated in the literature; for example, \citet{Valenti2005} neglect all stars colder than 4700~K from their abundance analysis ``due to significant blending by spectral lines with poorly known atomic and molecular line data.'' The downturn appears closer to 4800~K in our Hyades data set, although its location is likely metallicity dependent; increasing metallicity, like decreasing temperature, increases the complexity of the stellar spectrum. While many authors have appreciated this K-dwarf issue, it is nonetheless common to see works that apply similar techniques to low-temperature stars. We therefore stress: fitting theoretical models to high-resolution spectra of mid-K and cooler dwarfs likely results in underestimated metallicities. To facilitate estimation of the bias in \texttt{uberMS}/\texttt{SPC} metallicities as a function of $T_{\rm eff}$, we publish our polynomial fits to the Hyades data (Table~\ref{tab:leg}); however, we again caution that a fully appropriate correction likely has some dependence on metallicity, which is not probed by this sample.

Considering only the 32 stars with masses above 0.82M$_\odot$ ($T_{\rm eff} \gtrsim 4800$K), we can calculate the mean and sample standard deviation from each of the three methods. This yields a Hyades metallicity of [Fe/H]=$0.157\pm0.048$ dex for \texttt{uberMS} in TP mode, [Fe/H]=$0.149\pm0.035$ dex for \texttt{uberMS} in MS mode, and [M/H]=$0.093\pm0.034$ dex for \texttt{SPC}. All three methods show some systematic trends with effective temperature. For alpha enrichment, we measure [\textalpha/Fe]=$0.006\pm0.015$ dex using TP mode and [\textalpha/Fe]=$0.008\pm0.012$ dex using MS mode.

\subsection{Gaia Benchmark Sample}
\label{sec:gbs3}
The Gaia FGK Benchmark Sample (GBS) is a collection of stars with $T_{\rm eff}$ and $\log g$ derived through fundamental relations. These stars can therefore be used to evaluate the performance of methods that infer these parameters from spectroscopy. Here we compare our results with Version 3 of the catalog from \citet{Soubiran2024}, which greatly increases the sample size over Versions 1 and 2 \citep{BlancoCuaresma2014, Heiter2015, Jofre2014, Jofre2015, Jofre2017, Jofre2018, Hawkins2016}. In total, GBS3 consists of 200 stars with precisely measured angular diameters; we consider 85 that have been observed by TRES. Many of these targets are evolved stars since it is easier to measure their diameters, but 35 have a GBS3 log$g > 4.0$~dex. These dwarf stars span a GBS3-reported temperature range of $3997 \mathrm{K} < T_{\rm eff} < 6914 \rm{K}$ and a metallicity range of $-2.12 < \rm{[Fe/H]} < 0.38$, although note that the GBS3 metallicities are taken mainly from the PASTEL catalog \citep{Soubiran2016, Soubiran2022}, which collates spectroscopic measurements from the literature; i.e., these metallicities are not independent direct measurements from fundamental relations.

Comparisons between the GBS3 parameters and our analysis of TRES spectra using \texttt{uberMS} and \texttt{SPC} are given in Figure~\ref{fig:gbs3}. For each comparison, we calculate two values that summarize the agreement: the median absolute deviation (MAD) and the root mean square (RMS). The MAD is more robust against outliers than the RMS; in cases where the MAD is much lower than the RMS, it provides an indication of the typical agreement if one neglects some egregious outliers (although note that the MAD is expected to be lower than the RMS in general, and must be multiplied by 1.48 to create a quantity analogous to the standard deviation; we refer to this value as $\sigma$MAD, and it is shown alongside the RMS in each subplot).

For dwarf stars, the three methods are relatively similar in their agreement with the GBS3 fundamental temperatures. However, this agreement is much poorer than the nominal $50$~K uncertainties reported by \texttt{SPC}, ranging from 128--130K, 103--120K, and 111--143K for TP mode, MS mode, and \texttt{SPC}, respectively, depending on whether one adopts 1.48$\times$MAD or the RMS as the error. The scatter in $\log g$ is 0.17--0.19 dex, 0.09--0.11 dex, and 0.06--0.09 dex, respectively. This ranking is unsurprising: TP mode does not use any isochronal information and thus has a much poorer ability to constrain $\log g$. On the other hand, the treatment of the isochronal prior in the \texttt{SPC} fit favors dwarf-like $\log g$ values, likely contributing to its smaller scatter; however, \texttt{SPC}'s $\log g$ results for evolved stars are much poorer than even the TP mode estimates, perhaps because \texttt{SPC}'s isochronal prior is uninformative in this regime. $T_{\rm eff}$ for the \texttt{SPC} evolved stars is correspondingly biased due to the poor $\log g$ fits. Note that \texttt{SPC} was designed with the specific goal of automatically measuring metallicities of Kepler planet host stars, which were almost exclusively dwarfs, and hence it was never intended or tested for subgiants and other evolved stars. Its poor performance in this regime is therefore not unexpected.

In the dwarf metallicity plots, the RMS is greatly inflated over the MAD due to some erroneous measurements associated with low-temperature stars; this is the same phenomenon we discussed in reference to the Hyades. Neglecting these, both \texttt{uberMS} modes exhibit good agreement with the GBS3 values, with $\sigma$MADs of 0.045~dex (TP) and 0.041~dex (MS). These errors are comparable to the scatter we saw for the Hyades stars (although the evolved star metallicities disagree with GBS3 by a much larger amount, with $\sigma$MADs of 0.15~dex and 0.13~dex, respectively). These results perhaps indicate that the precision attainable by these methods in the systematics-limited regime is 0.04--0.05~dex for dwarf stars. However, we again note the caveat that the GBS3 metallicities are not independently determined; it may be that \texttt{uberMS} measurements are in agreement with PASTEL due to shared biases. PASTEL collates metallicity measurements from a variety of surveys, and many of these surveys use a similar spectral synthesis fitting framework. Notably, the coldest stars in this overlap sample (the four outliers that are the most egregious in the dwarf metallicity plots in Figure~\ref{fig:gbs3}) have PASTEL metallicities determined using a different technique. For those stars, the measurements that GBS3 quote were taken from \citet{Luck2005, Luck2006}/\citet{Luck2017}, who measure abundances by making equivalent width measurements of specific metal lines rather than fitting a entire wavelength range like \texttt{uberMS} and \texttt{SPC}. In regimes where there are significant discrepancies between the observed stars and theoretical models -- like in the cases of these cool dwarfs -- it is possible that such techniques yield better results than fitting the entire spectrum and diluting small metallicity-sensitive signals with large amounts of systematic noise.

As for \texttt{SPC}, the $\sigma$MAD for dwarf metallicity is 0.098~dex, considerably larger than \texttt{uberMS}. However, this larger error is primarily driven by the metallicities having a systematic offset rather than increased scatter, with the \texttt{SPC} estimates being lower than GBS3 on average with a median offset of 0.06 dex. If this median offset is subtracted before calculating the standard deviation, we find an error of 0.049~dex, in closer alignment with the \texttt{uberMS} values. Some of this systematic offset may arise due to the differences between measuring [M/H] and [Fe/H] (these quantities are only equivalent if stars have the same relative abundance pattern as the Sun, an assumption that does not hold when alpha enrichment is non-zero); other notable distinctions between the codes are their underlying model atmospheres (ATLAS-12 vs ATLAS-9) and their treatment of microturbulence (SPC assumes $v_{\rm mic}=2$kms$^{-1}$ for all stars; \citealt{Torres2012}).

\begin{figure}[t]
\centering
    \includegraphics[width=\columnwidth]{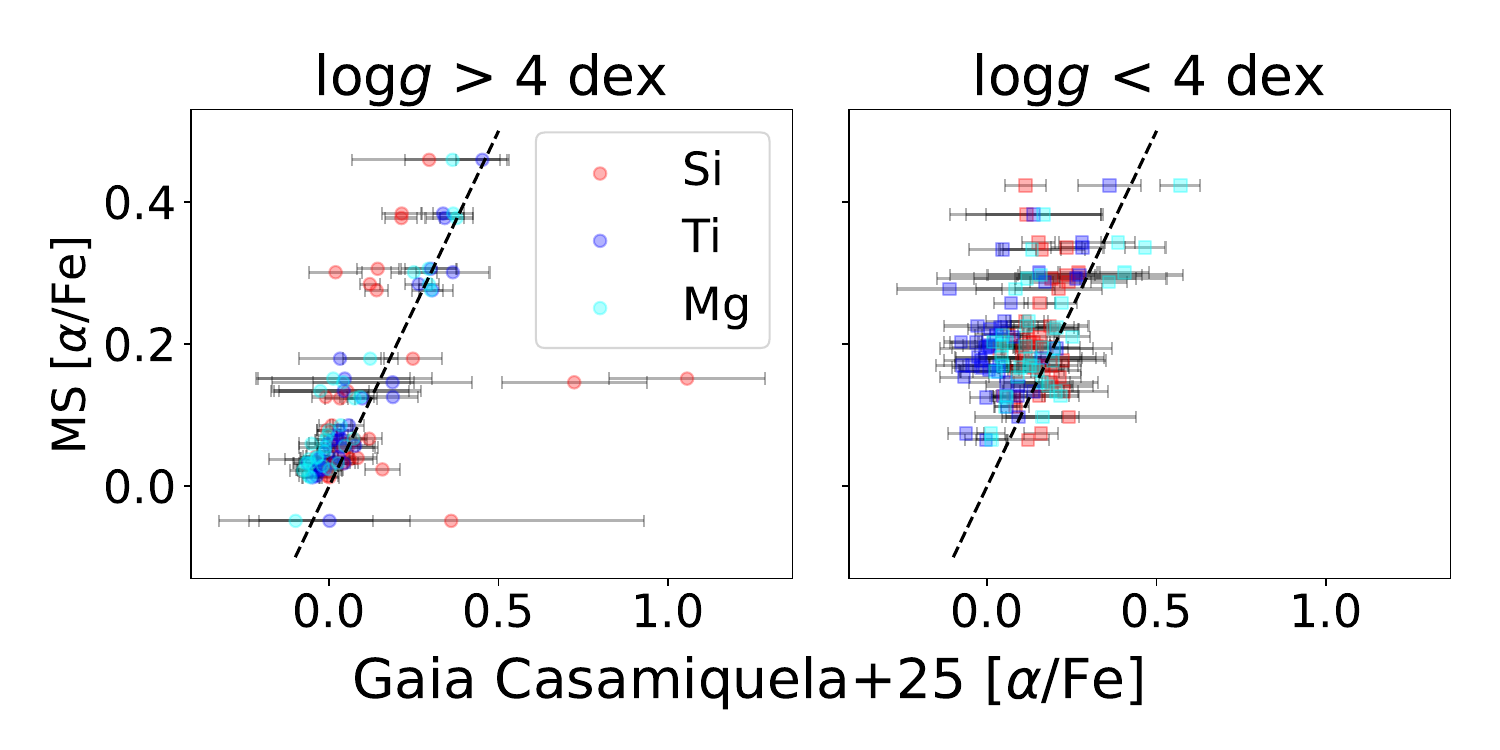}
    \includegraphics[width=\columnwidth]{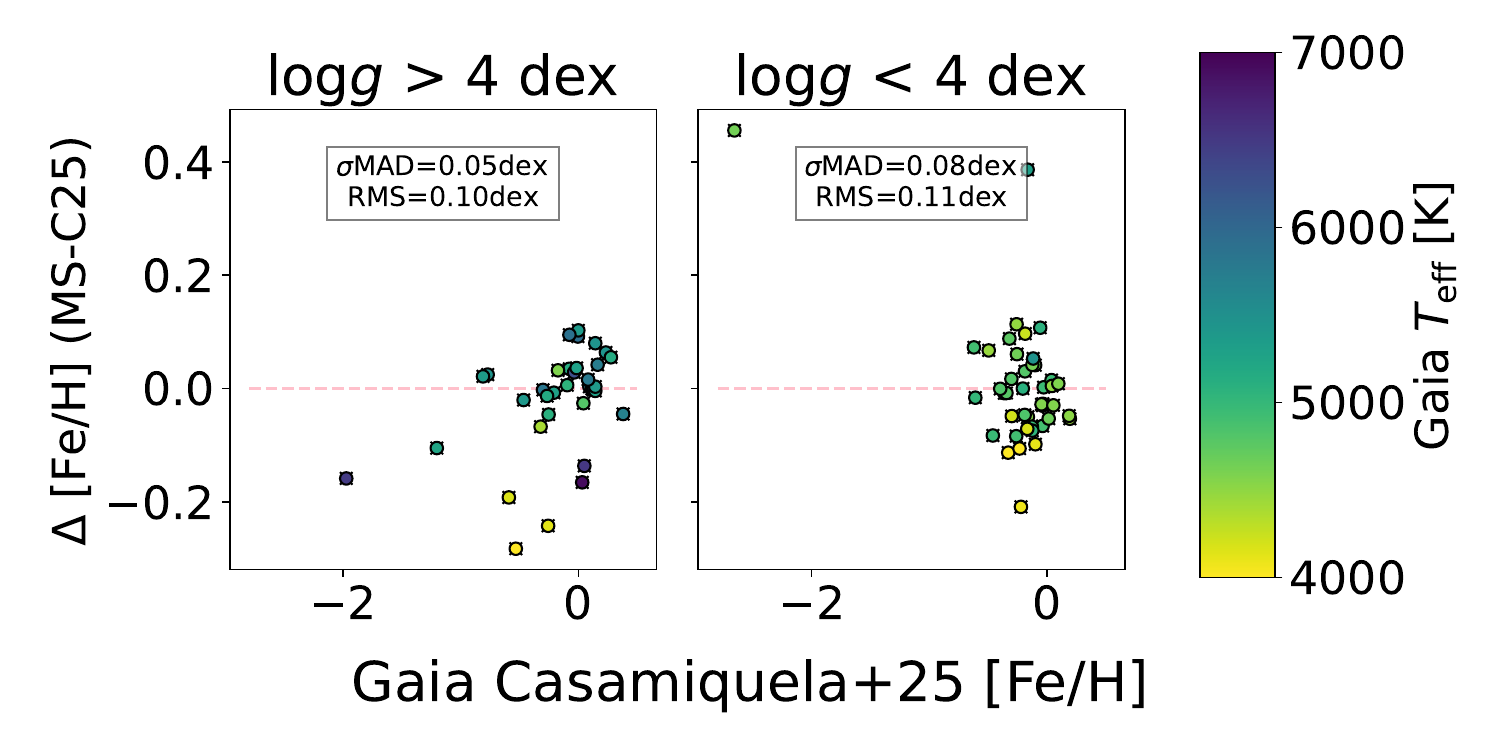}
    \caption{Our same \texttt{uberMS} analysis from panel B of Figure~\ref{fig:gbs3}, but compared to the iron abundance and alpha enrichments for the GBS3 sample measured in \citet{Casamiquela2025}. The dashed lines indicate unity.}
    \label{fig:casa}
    \end{figure}

While the metallicities in \citet{Soubiran2024} come from the PASTEL catalog, a recent follow-up paper performed a homogeneous analysis of the GBS3 sample \citep{Casamiquela2025}. Those authors used the interferometrically determined $T_{\rm eff}$ and $\log g$ from \citet{Soubiran2024}, then applied \texttt{iSpec} \citep{BlancoCuaresma2014b} and four different radiative transfer codes to determine chemical abundances from high-resolution spectra. In Figure~\ref{fig:casa}, we compare our [Fe/H] and [\textalpha/Fe] measurements to the recommended values from \citet{Casamiquela2025}. For dwarf stars, the agreement is marginally poorer between our \texttt{uberMS} results for [Fe/H] as compared to our previous comparison with the PASTEL catalog: while the previous MS-mode analysis yielded a $\sigma$MAD of 0.041~dex, the \citet{Casamiquela2025} comparison yields 0.054~dex. However, we see much improved agreement for the evolved stars: while our comparison with PASTEL yielded 0.13~dex, our \texttt{uberMS} measurements agree with \citet{Casamiquela2025} within 0.078~dex. We also note an improved agreement for the cooler dwarfs. These stars were dramatic outliers in Figure~\ref{fig:gbs3}, and while our [Fe/H] measurements remain low when compared to \citet{Casamiquela2025}, the difference is much more modest. However, this improved agreement is not necessarily favorable. As we have discussed in relation to the Hyades, \texttt{uberMS} appears to severely underestimate the metallicity of cool dwarfs due to limitations of the underlying stellar models; \citet{Casamiquela2025} similarly note that their coolest stars have a much higher dispersion when compared to the PASTEL catalog, and they see the greatest disagreement between their four tested radiative transfer methods at these low temperatures. That said, they adopt the measurements from one of the methods, \texttt{TURBOSPECTRUM} \citep{Plez2012}, that they argue is minimally affected by this bias.

\begin{figure*}[p]
\centering
    \includegraphics[width=0.8\textwidth]{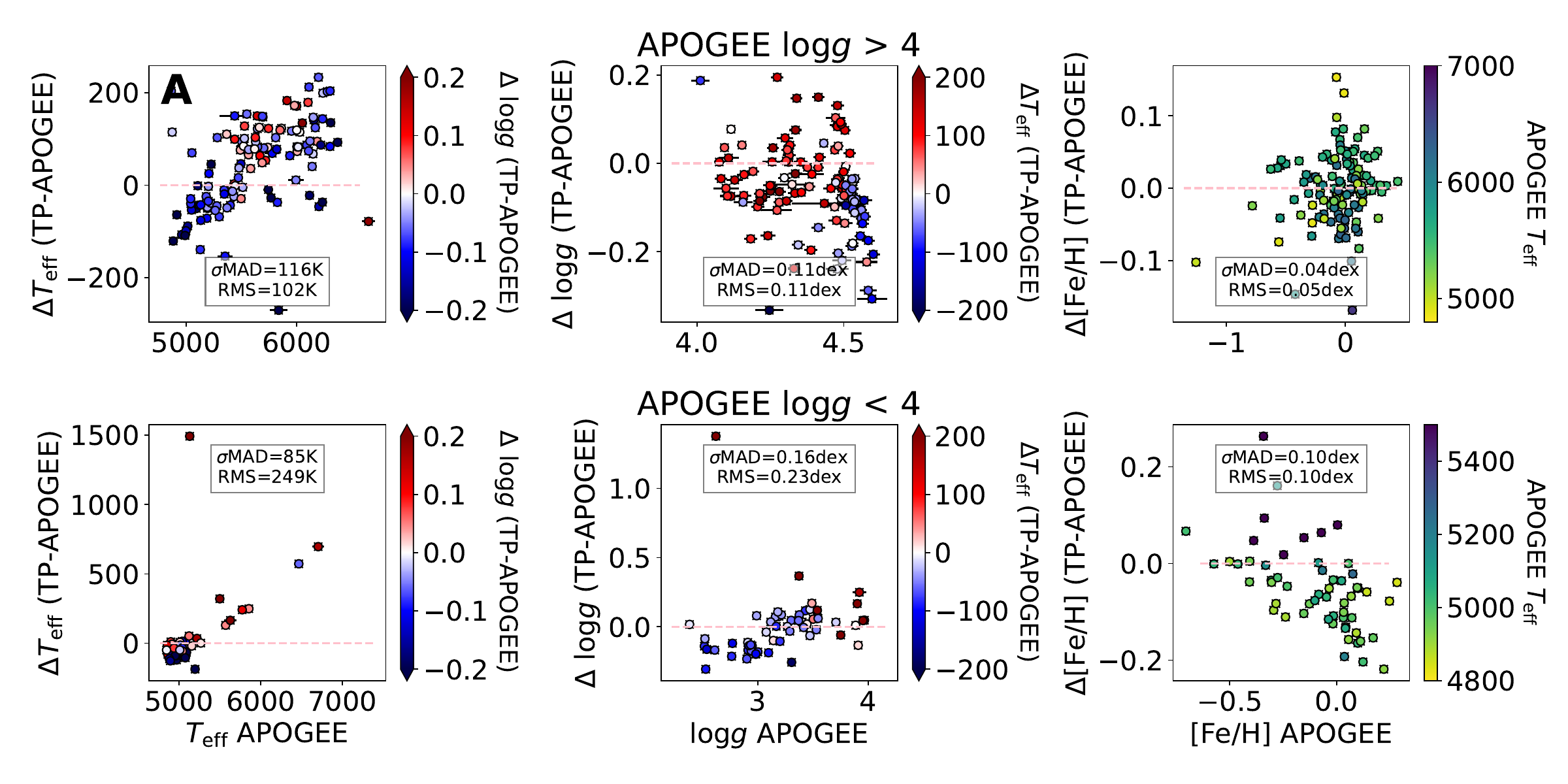}
    \includegraphics[width=0.8\textwidth]{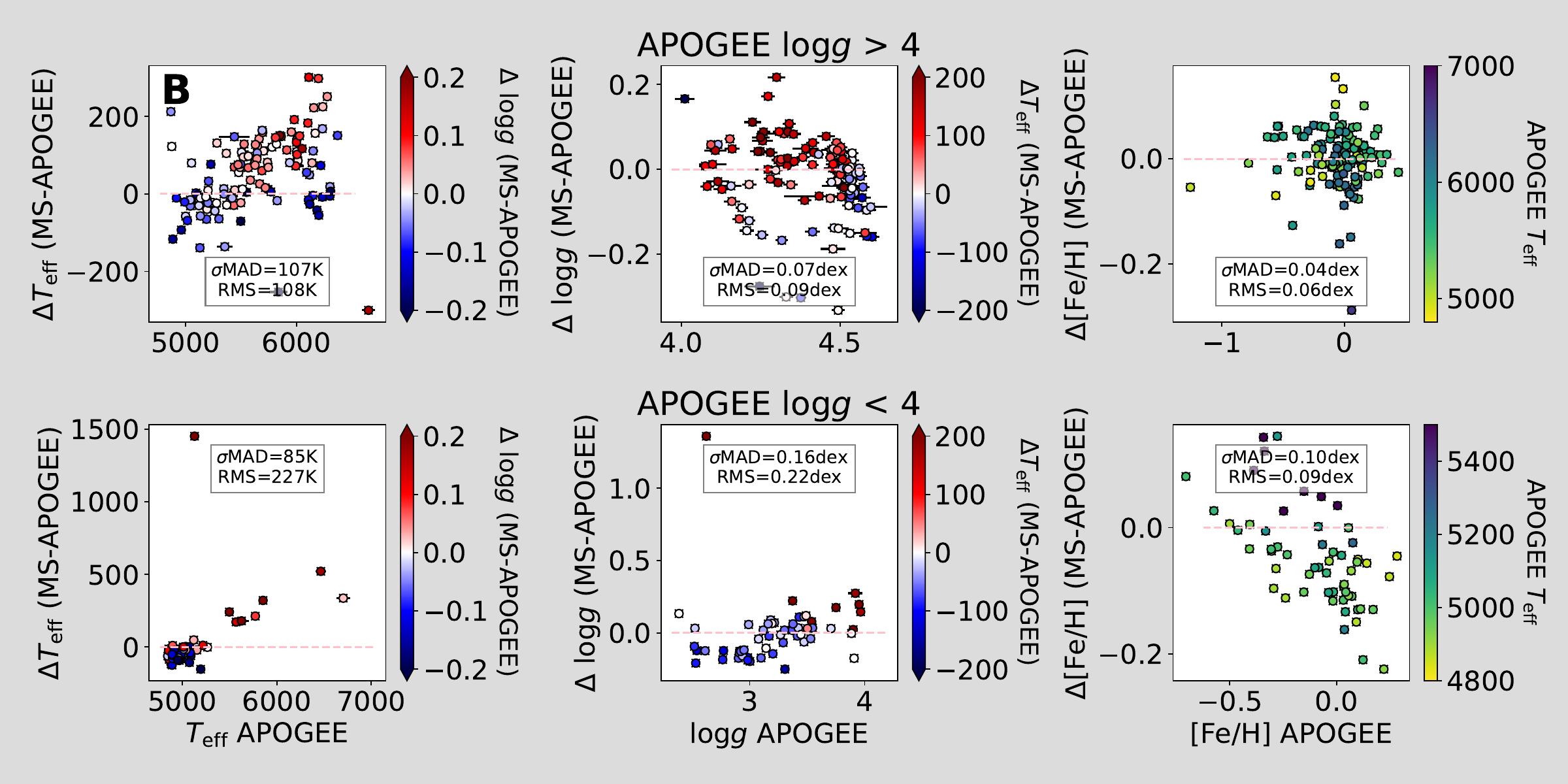}
    \includegraphics[width=0.8\textwidth]{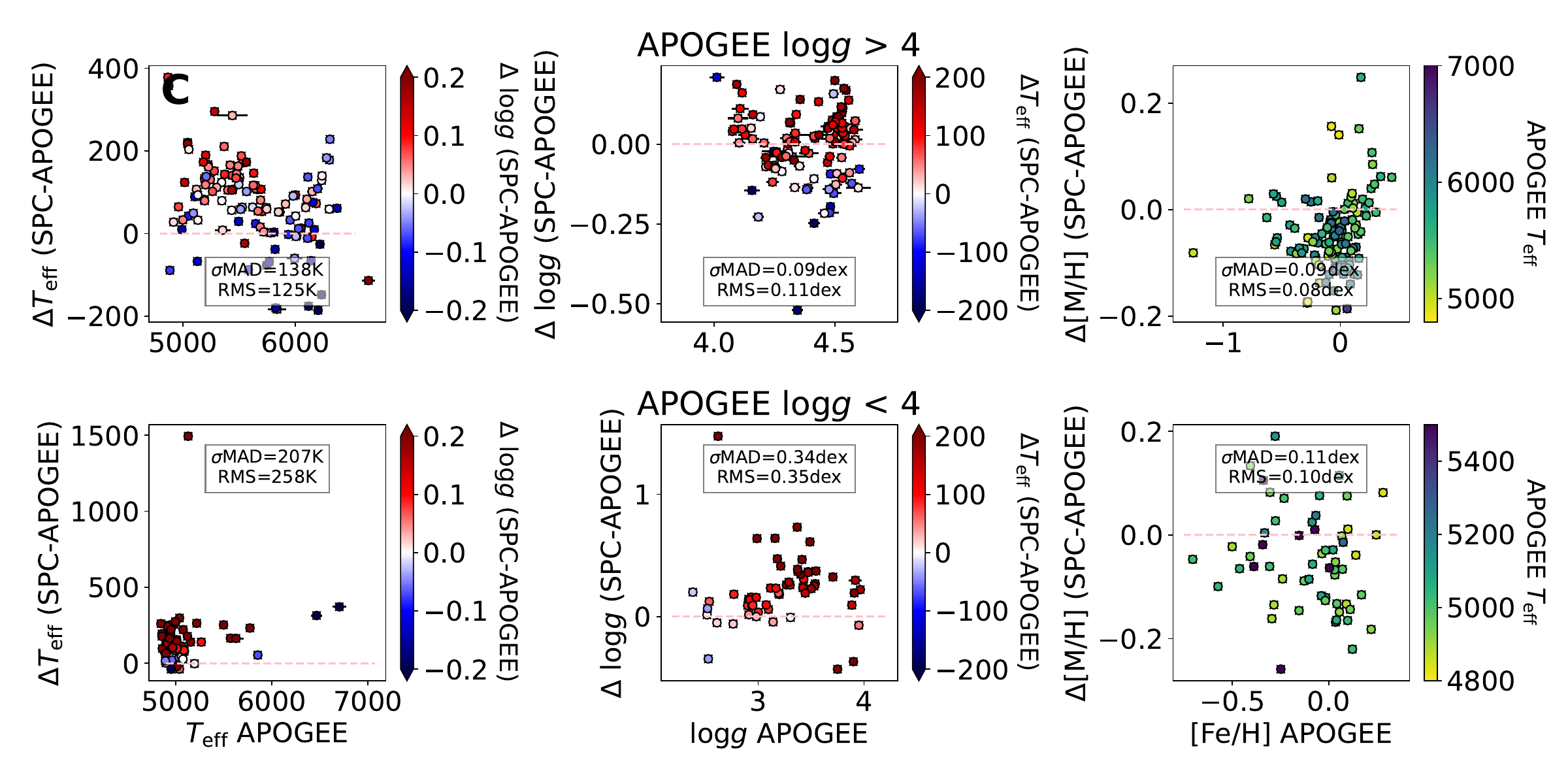}
    \caption{Comparison between APOGEE stellar parameters and properties estimated from TRES spectra using \texttt{uberMS} in TP mode (A), \texttt{uberMS} in MS mode (B), and \texttt{SPC} (C). In each set of subplots, the upper row shows dwarf stars (here approximated by log$g>4$ dex), while the lower row shows evolved stars (log$g<4$ dex). Note that error bars are only plotted for APOGEE values, not for our measurements. The dashed lines indicate unity.}
    \label{fig:apogee}
\end{figure*}

For alpha enrichment of the dwarfs, our comparison with \citet{Casamiquela2025} yields RMS errors of 0.057~dex for [Ti/Fe], 0.22~dex for [Si/Fe], and 0.074~dex for [Mg/Fe], or $\sigma$MADs of 0.072~dex, 0.11~dex, and 0.099~dex, respectively. There are systematic offsets between these various elements in \citet{Casamiquela2025}, with our [\textalpha/Fe] aligning best with their measurements of [Ti/Fe]. For the evolved stars, the scatter is 0.17--0.22~dex for [Ti/Fe], 0.097--0.10~dex for [Si/Fe], and 0.10--0.11~dex for [Mg/Fe], considering the two error estimation methods; the \citet{Casamiquela2025} elements again show offsets when compared to each other, with our [\textalpha/Fe] now aligning better with [Mg/Fe].

\subsection{APOGEE}
\label{sec:apogee}
In the previous section, we found a better agreement between our dwarf-star [Fe/H] estimates and the non-homogeneous PASTEL catalog than we did when comparing with the homogeneous analysis of the same stars from \citet{Casamiquela2025}. Such a result is surprising, perhaps implying that our method is more closely aligned with other techniques that are well represented in PASTEL. We may therefore gain further insight into our method's performance by comparing with other surveys. In this section, we make such a comparison with APOGEE; in Section~\ref{sec:spocs}, we compare with SPOCS.

The APO Galactic Evolution Experiment 2 \citep[APOGEE-2;][]{Majewski2017} is part of phase IV of the Sloan Digital Sky Surveys \citep{Blanton2017}, yielding $H$-band, $R$=22,500 spectroscopy of roughly 657,000 unique targets. Here we consider the stellar parameters determined using the ASPCAP pipeline \citep{GarciaPerez2016} in DR17, the final SDSS-IV data release \citep{Abdurrouf2022}.

To identify targets for our comparison sample, we begin with the full \texttt{allStarLite-dr17-synspec\_rev1.fits} catalog made available by the APOGEE team and downselect using a series of cuts: we reject stars flagged as bad in the \texttt{ASPCAPFLAG} variable, stars with large reported metallicity uncertainties (\texttt{FE\_H\_ERR} $>$ 0.05 or \texttt{MG\_FE\_ERR} $>$ 0.05), stars outside of our 4800--7000K range of interest, and stars that are unlikely to have been observed by TRES (either declination below $-25^{\circ}$, magnitude fainter than $G=10$~mag, or parallax less than 10~mas). After cross-matching with the TRES archive, we identify 177 targets suitable for our comparison. We show these results in Figure~\ref{fig:apogee}.

\begin{figure}[t]
\centering
    \includegraphics[width=\columnwidth]{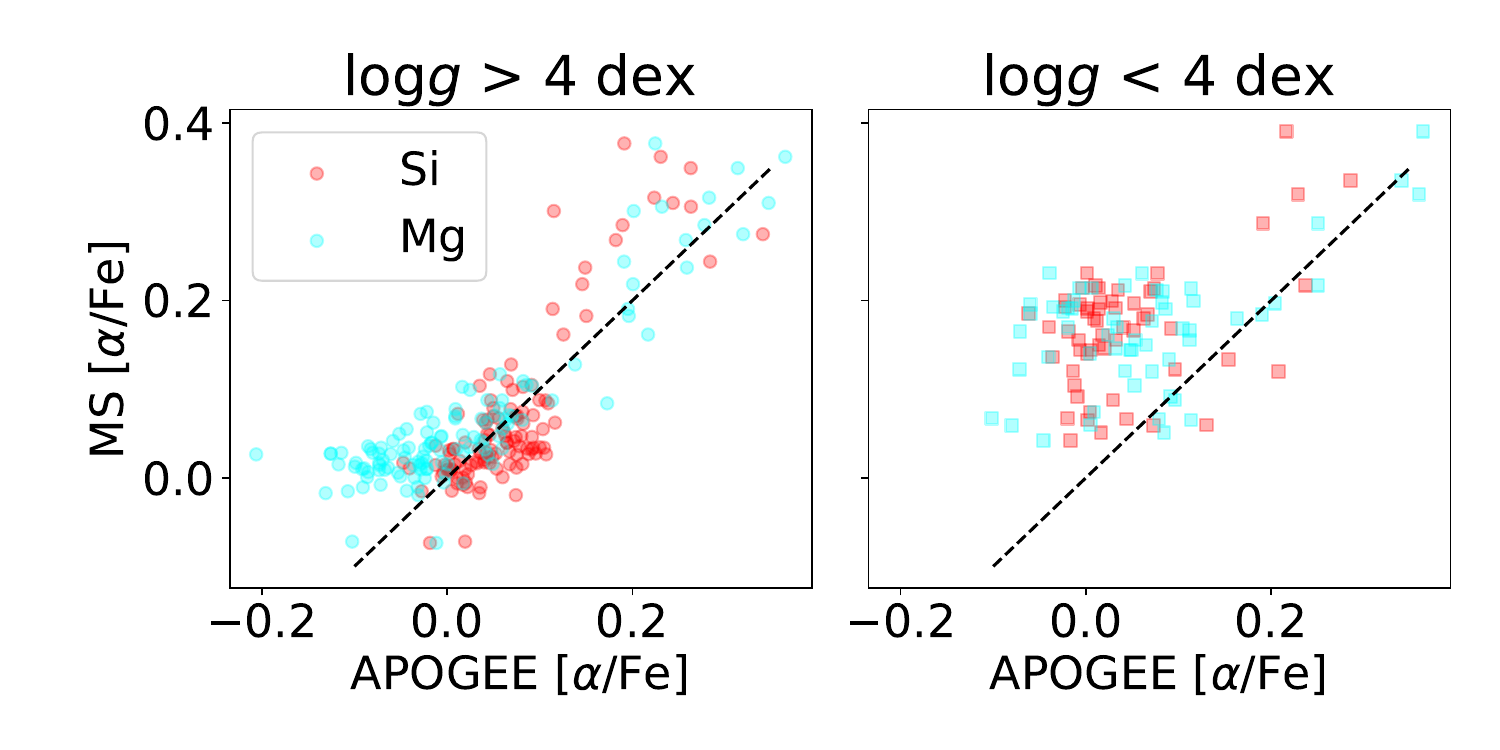}
    \includegraphics[width=\columnwidth]{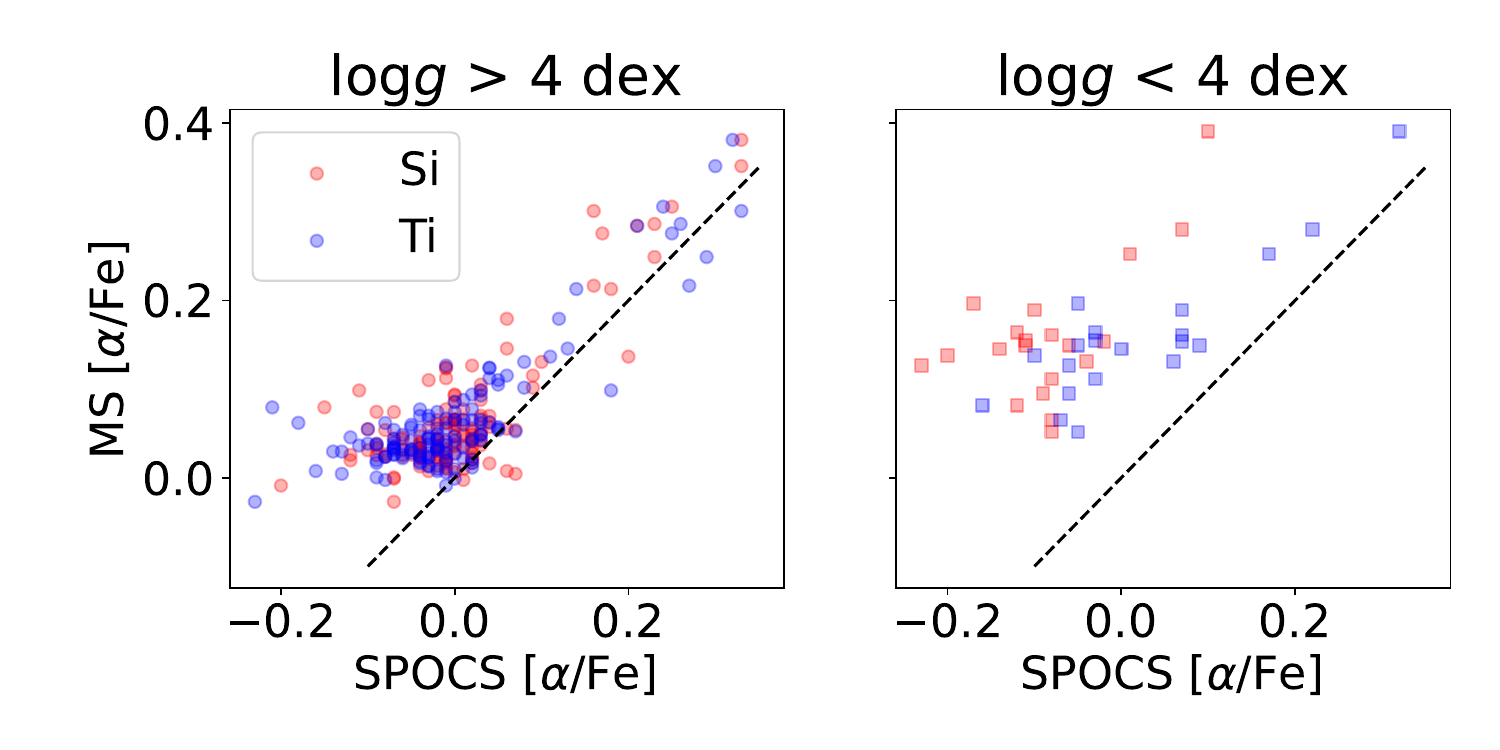}

    \caption{In the upper panels, we show comparison between APOGEE [Si/Fe] and [Mg/Fe] abundance ratios and the [\textalpha/Fe] ratio we estimate from \texttt{uberMS} in MS mode using TRES data. The lower panels compare SPOCS [Si/Fe] and [Ti/Fe] with our \texttt{uberMS} estimates; HD 82558 and HD 80133 have been neglected from this plot due to their potentially erroneous values (see text). The dashed lines indicate unity.}
    \label{fig:apogee_alpha}
\end{figure}

There is one striking outlier on these plots, appearing in the upper left in the $T_{\rm eff}$ and $\log g$ plots for the evolved stars in all three methods. APOGEE's results indicate that this star is cold and evolved, while \texttt{uberMS}/\texttt{SPC} return a hot dwarf. This target is Theta Cygni, which is known to be an F dwarf; the source of the discrepancy is therefore APOGEE, not our data analysis methods.

\begin{figure*}[!p]
\centering
    \includegraphics[width=0.8\textwidth]{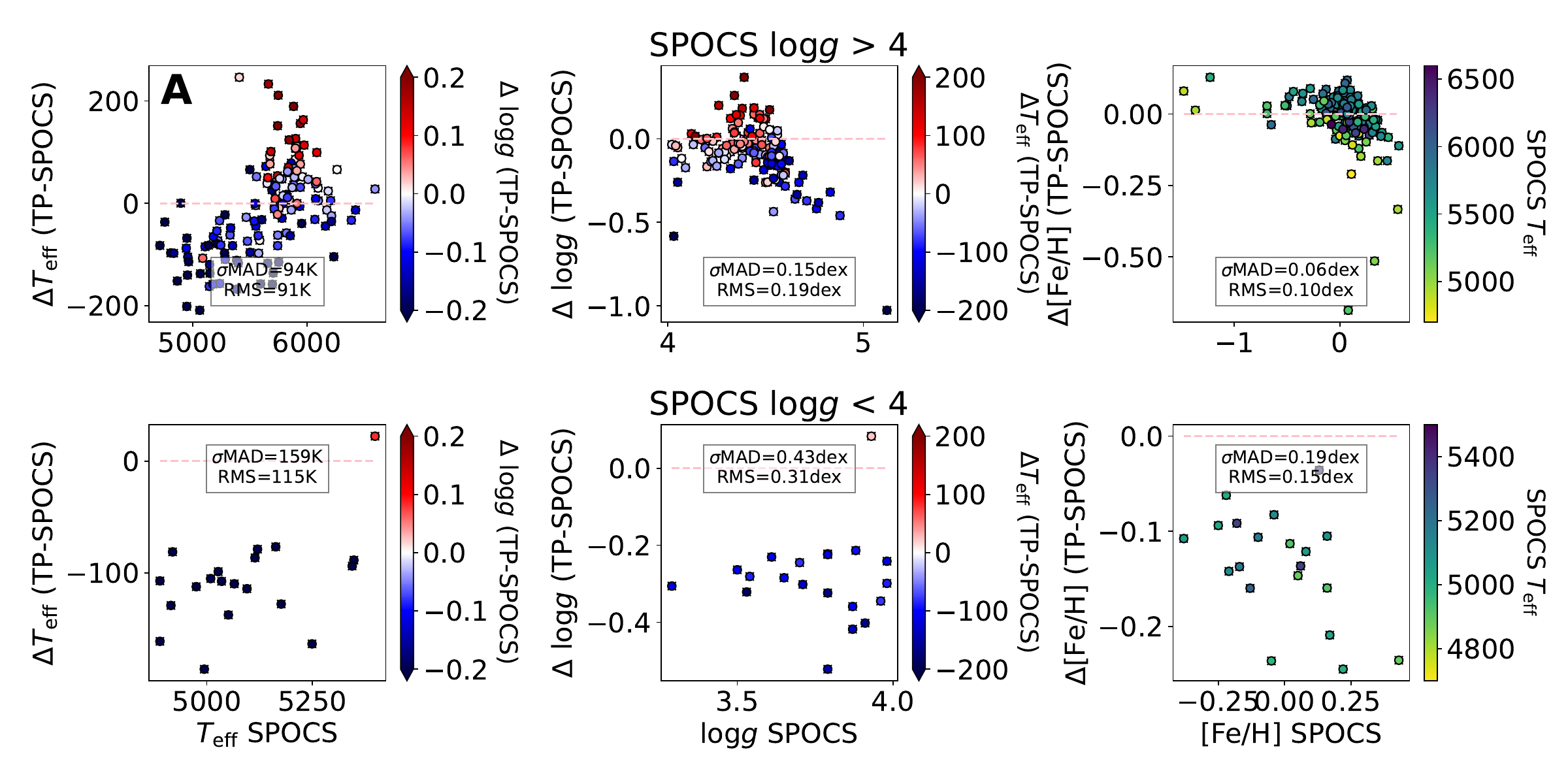}
    \includegraphics[width=0.8\textwidth]{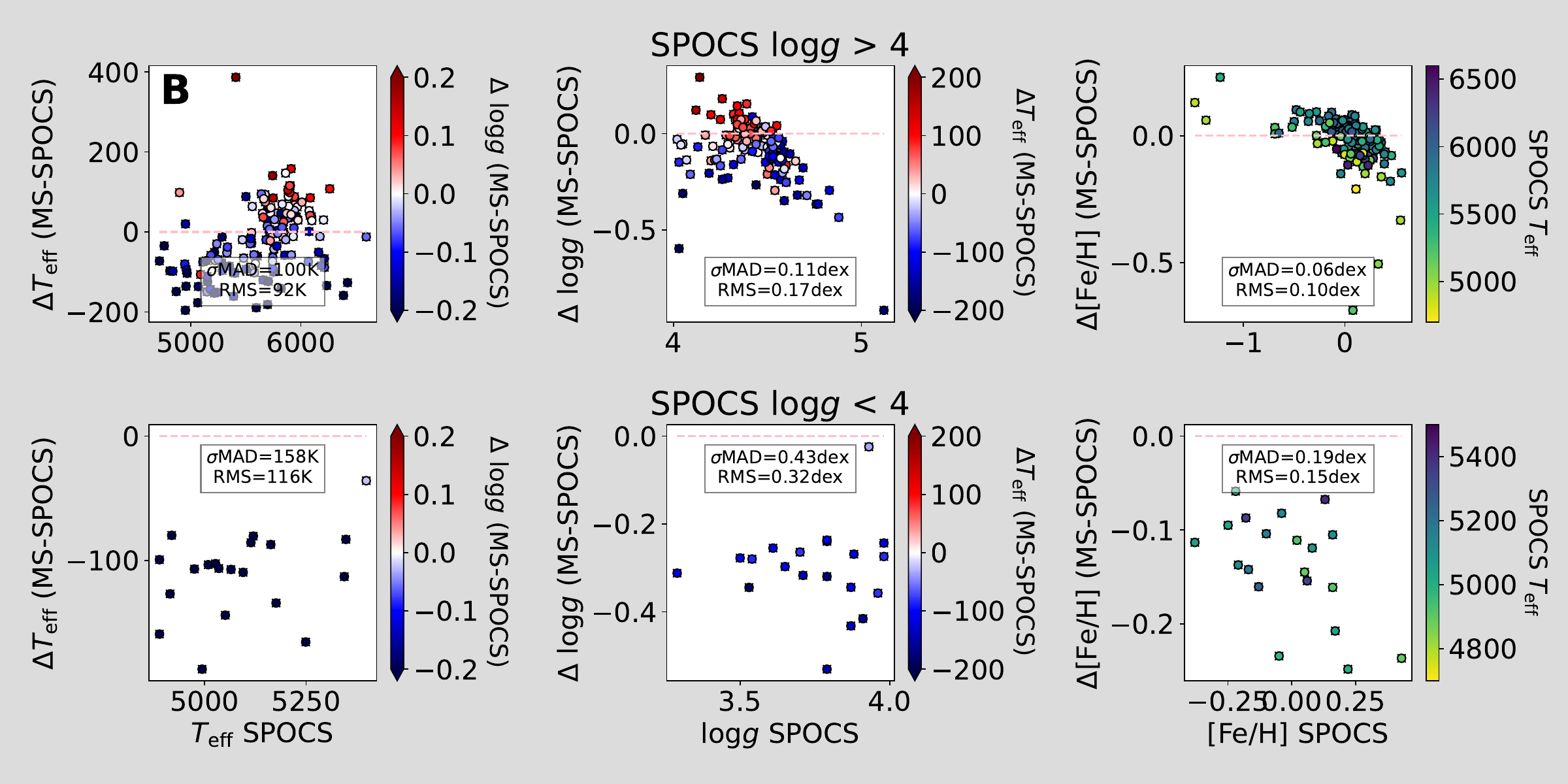}
    \includegraphics[width=0.8\textwidth]{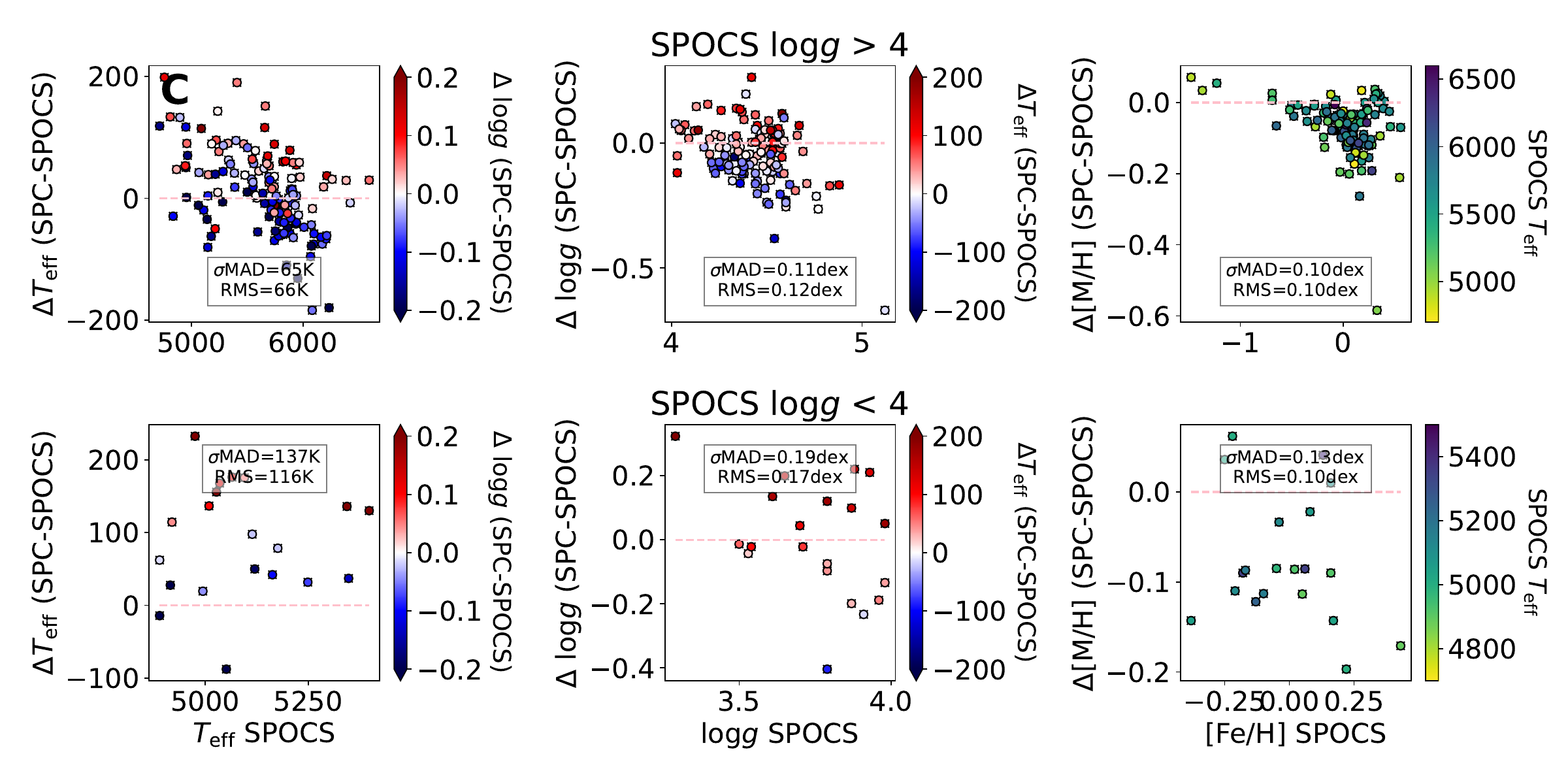}
    \caption{Comparison between SPOCS stellar parameters and properties estimated from TRES spectra using \texttt{uberMS} in TP mode (A), \texttt{uberMS} in MS mode (B), and \texttt{SPC} (C). In each set of subplots, the upper row shows dwarf stars (here approximated by log$g>4$ dex), while the lower row shows evolved stars (log$g<4$ dex). The dashed lines indicate unity.}
    \label{fig:spocs}
\end{figure*}

For dwarf stars, the metallicities returned by \texttt{uberMS} are in good agreement with APOGEE, with $\sigma$MADs of 0.041~dex (TP) and 0.042~dex (MS), or RMS errors of 0.048~dex and 0.056~dex, respectively. While the APOGEE measurements are subject to their own biases and systematics and cannot be taken as ground truth, recall that they represent $H$-band spectra; this technique is therefore more independent from \texttt{uberMS} and \texttt{SPC} than they are from each other, as those techniques both operate on optical spectra in the vicinity of the magnesium triplet. \texttt{SPC} yields a $\sigma$MAD of 0.091~dex and an RMS of 0.082~dex for [M/H], although as in the GBS3 comparison, the scatter is dominated by an offset of 0.051~dex. When that offset is subtracted, we find a $\sigma$MAD of 0.057~dex. The agreement is poorer for the evolved stars, where our three methods produce systematically lower metallicites than APOGEE.

In addition to bulk metallicities, APOGEE also provides measurements of alpha-element enrichments (e.g., [Mg/Fe], [Si/Fe], [Ti/Fe]), allowing us to compare these alpha enrichments with our predictions from \texttt{uberMS}. For MS mode, we show the results of this comparison in Figure~\ref{fig:apogee_alpha}; the results from TP mode are very similar. For evolved stars, the agreement is poor and the scatter is substantial. Agreement is better for the dwarfs: we find an RMS error of 0.069~dex for [Mg/Fe] and 0.050~dex for [Si/Fe], or a $\sigma$MAD of 0.074~dex and 0.039~dex. The \texttt{uberMS}-determined alpha enrichment therefore appears to align better with APOGEE's [Si/Fe] measurements than their [Mg/Fe] measurements, although for the small number of stars with large alpha enrichments, \texttt{uberMS} appears to predict larger values of this enrichment than APOGEE's [Si/Fe], which are in better agreement with their [Mg/Fe]. APOGEE does also provide estimates of [Ti/Fe], but we neglect them from this comparison due to their much larger errors.

\begin{figure*}[t]
\centering\includegraphics[width=0.8\textwidth]{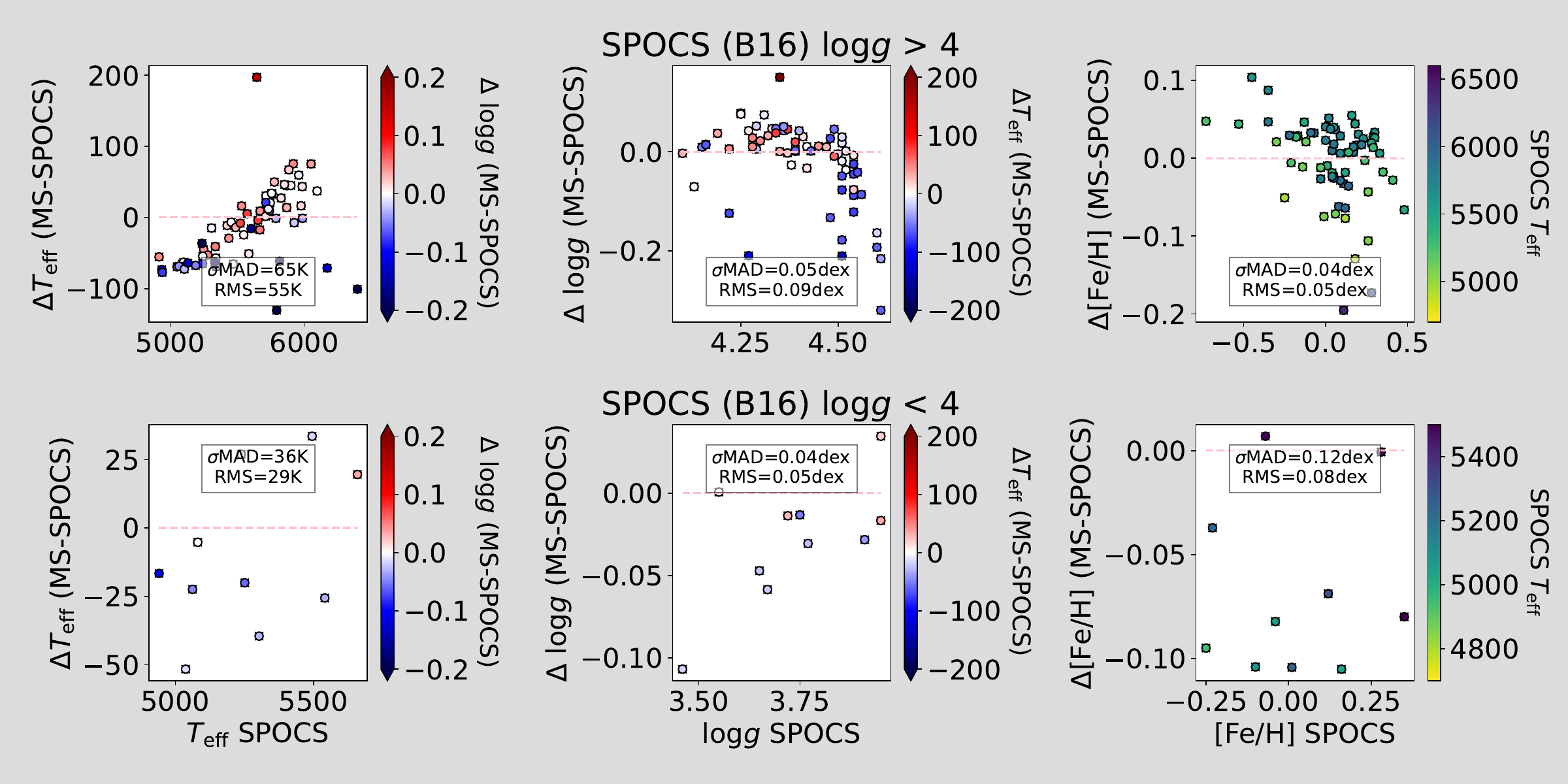}
    \caption{The same MS-mode \texttt{uberMS} vs SPOCS comparison as panel B of Figure~\ref{fig:spocs}, but replacing the \citet{Valenti2005} SPOCS parameters with the revised parameters from \citet{Brewer2016}. Some stars were not analyzed in \citet{Brewer2016} and hence are omitted. Dashed lines indicate unity.}
    \label{fig:brewer}
\end{figure*}

\subsection{SPOCS}
\label{sec:spocs}
As the intended application of this project is to measure the metallicities of planet-hosting stars, we also make a comparison with the SPOCS catalog \citep{Valenti2005}, which is commonly used in exoplanetary work. Specifically, this catalog uses the Spectroscopy Made Easy (\texttt{SME}) code \citep{Valenti1996} to measure bulk metallicity and individual elemental abundances from high-resolution optical spectra.
\begin{figure*}[t!]
\centering
    \includegraphics[width=\textwidth]{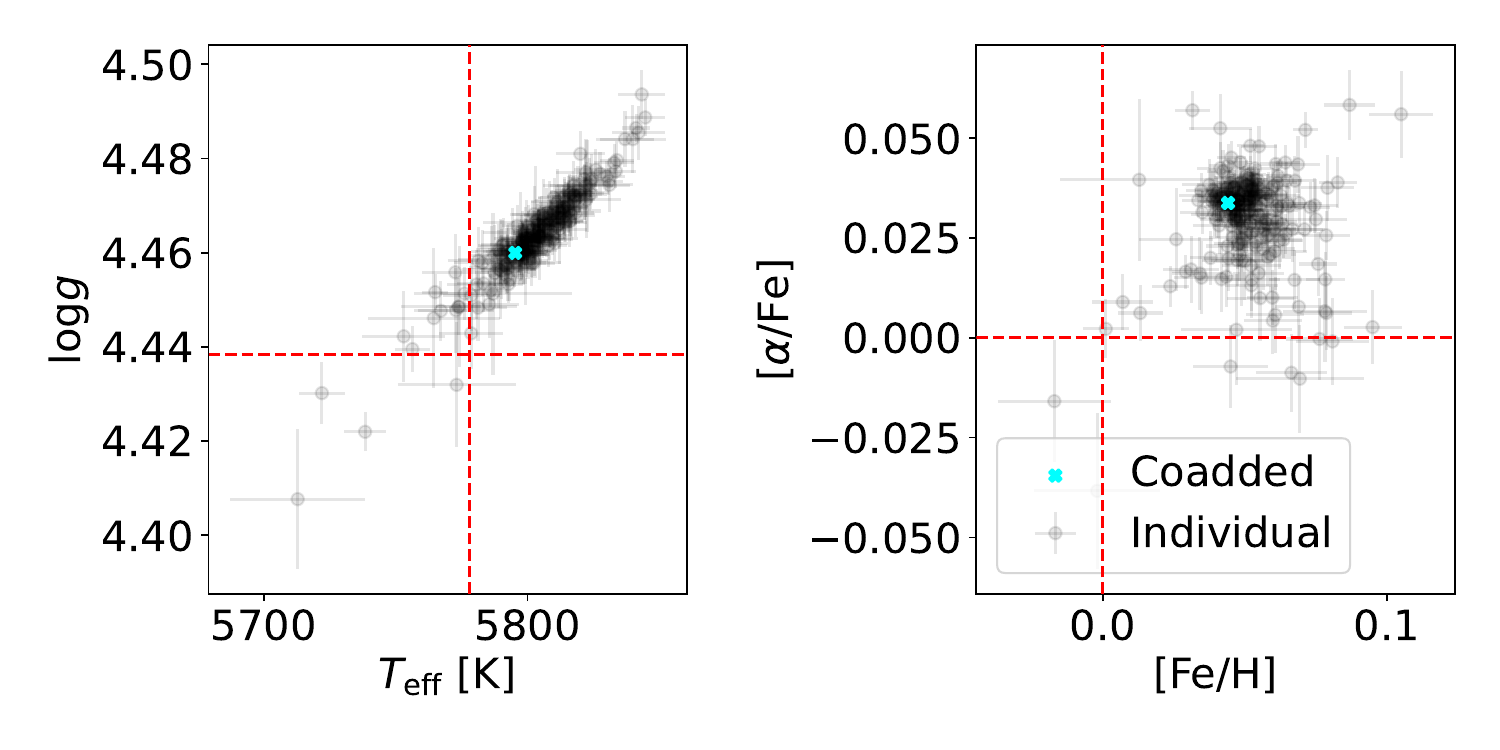}
    \caption{Results from \texttt{uberMS} in MS mode for TRES data of asteroids. Dashed lines indicate solar values. The black points represent each individual asteroid spectrum, while the cyan cross is a coadded spectrum of all observations of all asteroids.}
    \label{fig:asteroids}
\end{figure*}

Beginning with the 1039 stars in the \citet{Valenti2005} SPOCS catalog, we analyze a subset of 155 stars with data in the TRES archive. These results are shown in Figure~\ref{fig:spocs}. There are two obvious outliers in the dwarf metallicity plot: HD 82558 and HD 80133. In the case of HD 82558, \hbox{\citet{Valenti2005}} report highly discrepant bulk/iron metallicities of [M/H]=$-0.21$~dex and [Fe/H]=0.33~dex. We estimate [Fe/H]=$-0.18$~dex with \texttt{uberMS}, in reasonable agreement with the SPOCS-determined [M/H]. The discrepancy is therefore either an error in the SPOCS [Fe/H], or a case where the abundance pattern of individual elements is highly non-solar, which cannot be captured in our two-parameter ([Fe/H] and [\textalpha/Fe]) \texttt{uberMS} fit. For HD 80133, this star is a double-lined spectroscopic binary and so we expect all of \texttt{SME}, \texttt{uberMS}, and \texttt{SPC} to produce erroneous results.

Neglecting these two special cases, we find a $\sigma$MAD of 0.061~dex and an RMS scatter of 0.072~dex for [Fe/H] of dwarf stars with \texttt{uberMS} in MS mode, with comparable results for TP mode. The \texttt{SPC} metallicities are systematically lower than SPOCS with an offset of 0.066~dex, similar to what we saw in our previous comparisons with PASTEL and APOGEE. After subtracting that offset, the $\sigma$MAD is 0.051~dex and the RMS scatter is 0.089~dex. For evolved stars, all three methods produce metallicities that are substantially lower than SPOCS'; for \texttt{uberMS}, there are also similar offsets in $T_{\rm eff}$ and $\log g$.

In Figure~\ref{fig:apogee_alpha}, we also compare the alpha enrichments we measure from \texttt{uberMS} with those determined by SPOCS (specifically, as implied by the SPOCS elemental abundances of Si and Ti). For evolved stars, the correlation is generally poor. For dwarf stars, the techniques disagree on the value of alpha-enrichment for low-alpha stars; the \texttt{uberMS}-determined alpha enrichments range between 0~dex and 0.1~dex, while the SPOCS enrichments range from $-0.2$~dex to 0~dex. However, the techniques are in reasonable agreement regarding which stars are alpha-enriched: above 0.13~dex, the RMS error is 0.050~dex based on Ti and 0.074~dex based on Si. While this plot and analysis give values for MS mode, the performance of TP mode is similar.

\citet{Brewer2016} present an extension/revision of the SPOCS catalog, expanding the number of spectral lines included in the analysis to improve constraints on $T_{\rm eff}$ and $\log g$. In Figure~\ref{fig:brewer}, we repeat our \texttt{uberMS} analysis of the TRES--SPOCS overlap sample, replacing the SPOCS parameters from \citet{Valenti2005} with those from \citet{Brewer2016}. Of the 155 stars in our TRES--SPOCS overlap sample, 76 have revised parameters from \citet{Brewer2016}. The \citet{Brewer2016} analysis indeed appears to have improved the constraints on temperature and surface gravity: we find much better agreement with our \texttt{uberMS} results for both of these parameters, with $\sigma$MAD decreasing from 115~K to 54~K in $T_{\rm eff}$, 0.136~dex to 0.051~dex in $\log g$, and 0.061~dex to 0.048~dex in [Fe/H] when considering MS mode applied to the same subsample of 76 stars.

While the \citet{Brewer2016} results supersede those of \citet{Valenti2005}, our comparison with the original SPOCS catalog may still be relevant for some readers. For example, the \citet{Mann2013, Mann2014} relations, which are commonly used to estimate [Fe/H] of M and late-K dwarfs from spectroscopic indicators, are calibrated using \citet{Valenti2005} metallicities.

\subsection{Asteroids}
\label{sec:asteroids}
Asteroids are another handy calibrator for spectroscopic determinations of stellar properties, as they reflect the spectrum of our best-studied star, the Sun. We identify 234 asteroid spectra in the TRES archive, observed on various dates between 2011 and 2024. This collection includes observations of 21 different asteroids, with the largest contributions from Ceres, Eunomia, Hebe, Iris, Juno, Melpomene, Pallas, and Vesta.

From our \texttt{SPC} analysis of a high-SNR subsample of asteroid spectra, we measure typical values of $\mathrm{log}g=4.39$~dex, $T_{\rm eff}=5760$~K, and $\rm{[M/H]}=-0.060$~dex. The nominal uncertainties for \texttt{SPC} are 50~K in $T_{\rm eff}$, 0.1~dex in $\log g$, and 0.08~dex in [M/H]; our measurements are therefore consistent with the solar values within their reported errors, representing offsets of 18~K in temperature, 0.04~dex in $\log g$, and 0.06~dex in [M/H].

Although it can be run with spectroscopy alone, the \texttt{uberMS} method requires both spectroscopy and photometry to be performed in a manner consistent with the other analyses we have presented in this manuscript. For Gaia DR3, we adopt the solar absolute magnitude estimates from \citet{Creevey2023}: $M_{G,\odot}=4.66$, $M_{BP,\odot}=4.984$, and $M_{RP,\odot}=4.166$, and for 2MASS, we use the magnitudes tabulated in \citet{Willmer2018}: $M_{J,\odot}=3.67$, $M_{H,\odot}=3.32$, and $M_{K,\odot}=3.27$, which they determine from \citet{Cohen2003}.

\begin{figure*}[t]
\centering\includegraphics[width=0.75\textwidth]{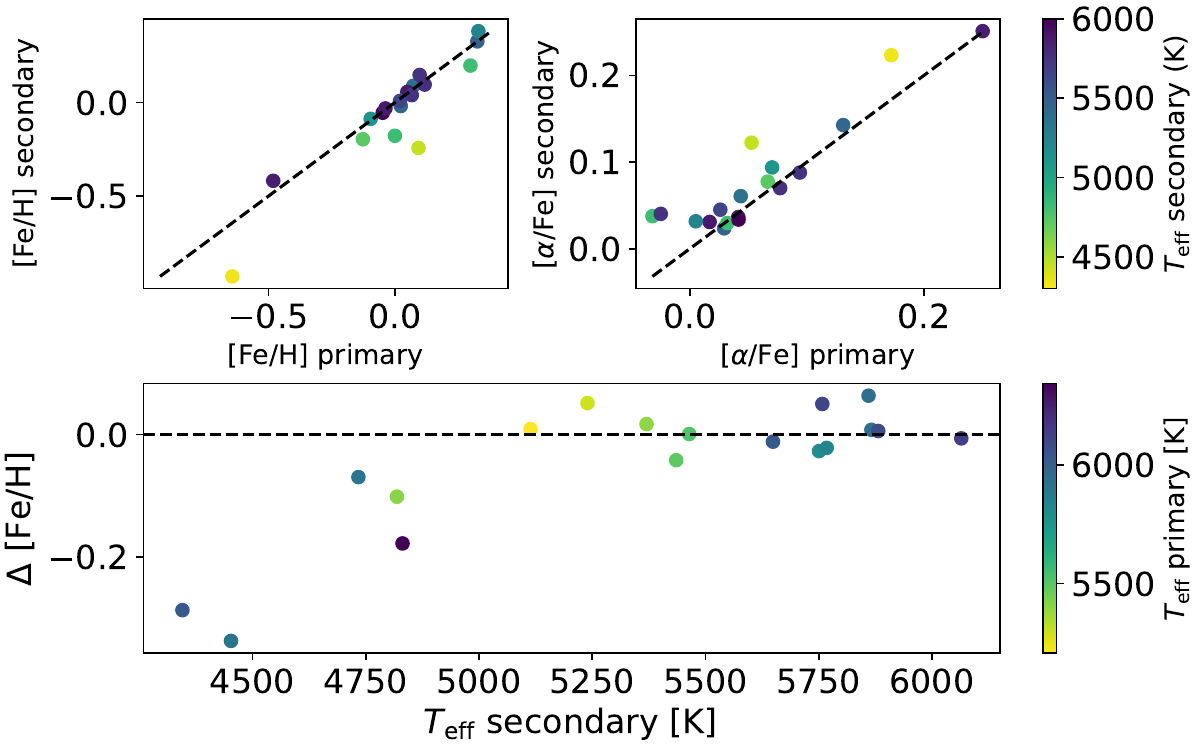}
    \caption{\texttt{uberMS} in MS mode applied to TRES data of the components of eighteen wide binaries. Dashed lines indicate unity. The upper panels compare the [Fe/H] (left) and [\textalpha/Fe] (right) estimates for the primary and secondary. The lower panel shows how the difference in [Fe/H] estimates varies with the effective temperature of the secondary: we observe a similar systematic offset as was seen in the Hyades data, where \texttt{uberMS} underestimates [Fe/H] for mid-to-late K dwarfs.}
    \label{fig:binaries}
\end{figure*}

Our \texttt{uberMS} MS-mode asteroid analysis is presented in Figure~\ref{fig:asteroids}, showing both the results for each asteroid spectrum analyzed individually as well as the coaddition of all 234 spectra; we do not observe any difference in parameters based on whether we average the asteroid spectra before or after the \texttt{uberMS} analysis. The difference between our measurement and the nominal solar value is 17~K in $T_{\rm eff}$, 0.022~dex in $\log g$, 0.044~dex in [Fe/H], and 0.034~dex in [\textalpha/Fe]. We also reanalyze our coadded asteroid spectrum using \texttt{uberMS}'s spectroscopy-only mode, finding that this changes our results relative to the previous fit by only 4~K in temperature, 0.004~dex in $\log g$, 0.0004~dex in [Fe/H], and 0.0007~dex in [\textalpha/Fe]. Our assumptions about solar magnitudes therefore have a negligible effect on the conclusions of this analysis. Reanalyzing the coadded spectrum with TP mode (including both spectroscopy and photometry, but lacking the isochronal information that is used in MS mode), our results differ by 1.5~K in temperature, 0.013~dex in $\log g$, 0.0005~dex in [Fe/H], and 0.002~dex in [\textalpha/Fe]; the two analyses are therefore highly similar.

While \texttt{uberMS} achieves good agreement with the expected solar effective temperature and surface gravity, the offset in metallicity is somewhat large, although consistent with our previous assertion based on the comparisons made in this work that the TRES--\texttt{uberMS} accuracy limit may be 0.04–0.05 dex for dwarf stars. While one might have expected that \texttt{uberMS} would achieve the best performance near solar values, as various aspects of the stellar models are informed by our knowledge of the Sun, this asteroid analysis suggests that solar twins are not immune to the systematic errors intrinsic to these fitting methods. In particular, note that we saw a similar +0.05~dex offset for Hyades stars at Sun-like temperatures relative to our Hyades average ([Fe/H]=0.149~dex for $M_*>0.82$M$_\odot$), suggesting that the offset in the asteroids is a temperature-dependent systematic; ergo, we cannot correct this bias by simply subtracting 0.044~dex from all our \texttt{uberMS} [Fe/H] measurements. It is likely more appropriate to subtract a temperature-dependent correction using the fit from Table~\ref{tab:leg} (such a treatment would yield [Fe/H]=0.008~dex for our asteroid sample), although we lack the data to determine whether this correction is valid for lower metallicities.

\begin{table}[t]
\centering
\caption{Binary pairs with individual TRES spectra}
\label{tab:binaries}
\begin{tabular}{@{}ll@{}}
\toprule
\textbf{Component A} & \textbf{Component B} \\ \midrule
16~Cyg~A & 16~Cyg~B \\
EPIC~201357835 & EPIC~201357643  \\
HD~13043~A & HD~13043~B \\
HD~19055~A & HD~19055~B \\
HD~26735~A & HD~26735~B \\
HD~26923 & HD~26913 \\
HD~30101~A & HD~30101~B \\
HD~35961~A & HD~35961~B \\
HD~59984~A & HD~59984~B \\
HD~80606 & HD~80607 \\
HD~99491 & HD~99492 \\
HD~179958 & HD~179957 \\
HD~190042~A & HD~190042~B \\
Kepler-25 & KOI-1803 \\
Weis 50670 & Weis 50671 \\
Weis 23083 & Weis 23082 \\
Weis 43227 & Weis 43228 \\
XO-2~S & XO-2~N \\ \bottomrule
\end{tabular}
\end{table}

\subsection{Wide binaries}
\label{sec:binaries}
Wide binaries provide a similar test to our method as open clusters, having formed together and thus sharing a common age and metallicity (neglecting the effects of diffusion). We perform a common-proper motion search for binaries and search the literature (and in particular, the Gaia wide binary catalog of \citealt{elBadry2021}), identifying 18 widely separated dwarf star pairs with TRES observations (Table~\ref{tab:binaries}). We neglect evolved stars and stars with known/suspected subarcsecond companions, as such properties can lead to larger errors.

Figure~\ref{fig:binaries} shows the results of our \texttt{uberMS} MS-mode analysis of the binary pairs. In general, we find good agreement between the metallicities we determine for the components; if we consider only stars hotter than 5000~K, we find that [Fe/H] has an RMS of 0.031~dex and a $\sigma$MAD of 0.025~dex, while [\textalpha/Fe] has an RMS of 0.023~dex and a $\sigma$MAD of 0.017~dex. The results for TP mode are similar (0.038~dex, 0.034~dex, 0.022~dex, and 0.012~dex, respectively), while \texttt{SPC} achieves an RMS of 0.040~dex and a $\sigma$MAD of 0.049~dex for agreement in [M/H]. However, these methods perform poorly for colder stars: as Figure~\ref{fig:binaries} illustrates, \texttt{uberMS} yields an offset of roughly $-0.1$~dex at 4800~K and $-0.3$~dex at 4400~K. This is similar to the behavior we observed in the Hyades, again highlighting the unreliability of spectroscopically determined [Fe/H] at low temperatures. Note that this trend cannot be explained as a signature of diffusion, as diffusion would cause a bias in the opposite direction: on the main sequence, the surface metallicity decreases over time as metals sink, but this process takes longer for later-type stars with deeper surface convection zones \citep[e.g., Figure 4 of][]{Dotter2017}.

\section{Summary}
\label{sec:summary}
In this work, we apply the \texttt{uberMS} method to archival TRES spectra in order to benchmark its performance for determining [Fe/H] and [\textalpha/Fe] of nearby stars, with the goal of establishing appropriate uncertainties to adopt when employing this technique in various exoplanetary applications. We analyze each data set with both of \texttt{uberMS}'s two modes: TP mode, which includes only spectroscopy and photometry, and MS mode, which includes spectroscopy, photometry, and isochrones. We also perform a complementary analysis with \texttt{SPC} \citep{Buchhave2012}, another fitting method that is often used for TRES data but which lacks the ability to estimate alpha enrichment. Our tests include comparisons to stars with external constraints on metallicity (the Hyades cluster, asteroids, wide binaries), as well as to other works that have estimated [Fe/H] and [\textalpha/Fe] using different fitting methods (SPOCS and APOGEE); we also compare with the Gaia Benchmark Sample (GBS3) of stars with interferometric radii, which provides external constraints on $T_{\rm eff}$ and $\log g$ but which still requires model fitting to obtain abundances.

\texttt{uberMS} in MS mode typically outperforms TP mode and \texttt{SPC}, with specific performance metrics summarized in Table~\ref{tab:results}. Synthesizing the various tests described in this manuscript, we find that TRES--\texttt{uberMS} (MS mode) provides [Fe/H] estimates with errors of roughly 0.04~dex for nearby dwarf stars, at least within the 5000--6200~K temperature regime. Performance is significantly poorer for mid-to-late K dwarfs, with this bias worsening as effective temperature decreases. The bias may depend on stellar properties and therefore further investigation is warranted to determine an appropriate correction; that said, our results suggest that the offset is roughly 0.1~dex at 4800~K and 0.3~dex at 4400~K. It is possible that the \texttt{uberMS} method may also yield larger errors for $T_{\rm eff} > 6200$~K or [Fe/H]~$<-1$~dex; as our comparison samples contain few stars in these regimes, we are unable to quantify such uncertainties.

From comparison with the GBS3 sample of stars with interferometric radii, we find that TRES--\texttt{uberMS} (MS mode) achieves errors of roughly 100~K in $T_{\rm eff}$ and 0.09~dex in $\log g$ for dwarfs. The error in $T_{\rm eff}$ is comparable for evolved stars, but the error in $\log g$ grows to 0.14~dex. Assessing the accuracy of the \texttt{uberMS} [Fe/H] estimates for evolved stars is difficult since we lack a comparison sample with external constraints; we note that the \texttt{uberMS} values are systematically lower than those from GBS3--PASTEL, APOGEE, and SPOCS (both \citealt{Valenti2005} and \citealt{Brewer2016}), although there is no offset when compared with the revised GBS3 metallicities from \citet{Casamiquela2025}. For [\textalpha/Fe], our agreement with other surveys depends on which \textalpha\ element is used to infer the enrichment; in general, the literature comparisons show systematic differences between [Si/Fe], [Mg/Fe], and/or [Ti/Fe]. Based on our asteroid, wide-binary, and Hyades samples, our [\textalpha/Fe] error may be as good as 0.03~dex for dwarf stars; however, these samples with external constraints all probe thin-disk stars with a relatively small dynamic range in alpha enrichment, and so the true error may be larger when the full diversity of compositions is considered.

TRES--\texttt{uberMS} is therefore capable of measuring precise [Fe/H] and [\textalpha/Fe] values for nearby stars, albeit with some limitations. Our analysis lays the groundwork for future investigations that apply the \texttt{uberMS} technique to the rich archival dataset available from TRES, as well as to new targeted observations, with upcoming studies from our team reporting its application to TESS planet demographics (PI: Rodr\'iguez Mart\'inez), thick-disk hot Jupiters (PI: DiTomasso), and M-dwarf planets through calibration with FGK-M wide binaries (PI: Pass).

\section*{Acknowledgments}
E.P.\ is supported by a Juan Carlos Torres Postdoctoral Fellowship at the Massachusetts Institute of Technology, V.D.\ acknowledges support from the National Science Foundation Graduate Research Fellowship under Grant No.\ DGE1745303, R.R.M.\ is supported by a Harvard Postdoctoral Fellowship for Future Faculty Leaders, and A.V.\ is supported in part by a Sloan Research Fellowship.

The authors are grateful to the many people who contributed to the development and operation of the TRES instrument over the past two decades, including Perry Berlind, Michael Calkins, Gilbert Esquerdo, Pascal Fortin, Jessica Mink, and Andrew Szentgyorgyi. We also thank Charlie Conroy for helpful conversations. The core neural-network concepts of the \texttt{uberMS} code can be traced back to the original \texttt{The Payne} algorithm, which was jointly developed by Yuan-Sen Ting and P.C.

This work used data from the European Space Agency (ESA) mission Gaia (\url{https://www.cosmos.esa.int/gaia}), processed by the Gaia Data Processing and Analysis Consortium (DPAC, \url{https://www.cosmos.esa.int/web/gaia/dpac/consortium}). Funding for the DPAC is provided by national institutions, in particular the institutions participating in the Gaia Multilateral Agreement. We also used data products from the Two Micron All Sky Survey, which is a joint project of the University of Massachusetts and the Infrared Processing and Analysis Center/California Institute of Technology, funded by the National Aeronautics and Space Administration and the National Science Foundation.

%

\facilities{FLWO:1.5m (TRES), Gaia}
\software{\texttt{Astropy} \citep{Astropy2022}, \texttt{JAX} \citep{Bradbury2018}, \texttt{Matplotlib} \citep{Hunter2007}, \texttt{NumPy} \citep{Harris2020}, \texttt{NumPyro} \citep{Phan2019}, \texttt{pandas} \citep{Reback2021}, \texttt{SciPy} \citep{Scipy2020}}


\appendix
\section{Long Tables}
\restartappendixnumbering

\begin{deluxetable*}{lllcc}[h]
\label{tab:priors}
\tabletypesize{\footnotesize}
\tablecolumns{7}
\tablewidth{0pt}
 \tablecaption{Fitting parameters and priors used with \texttt{uberMS}}
 \tablehead{
 \colhead{Parameter} &
 \colhead{Variable} &
 \colhead{Unit} &
 \colhead{Initial Value} &
 \colhead{Prior}}
\startdata
\multicolumn{5}{c}{\emph{Spectroscopic parameters}} \\
Resolution (per order) & lsf\_\{ii\} & --- & 44000 & $\mathcal{N}(44000, 1000)$; truncated at 40000 and 53000\\
Spectroscopic jitter (per order) & specjitter\_\{ii\} & --- & 0.015 & fixed \\
Radial velocity (per order) & vrad\_\{ii\} & kms$^{-1}$ & 0.0 & $\mathcal{U}(-5, 5)$\\
Normalization (per order) & pc0\_\{ii\} & --- & 1.0 & $\mathcal{N}(1.0, 0.01)$; truncated at 0.75 and 2.0 \\
Linear trend (per order) & pc1\_\{ii\} & --- & 0.0 & $\mathcal{U}(-0.2, 0.2)$ \\
Stellar broadening & vstar & kms$^{-1}$ & 2.0 & $\mathcal{U}(0, 250)$\\
Microturbulence & vmic & kms$^{-1}$ & Bruntt+2012 & Bruntt+2012 empirical relation\\
\\
\multicolumn{5}{c}{\emph{Photometric parameters}} \\
Extinction & Av & mag & 0.01 & $\mathcal{N}(0., 0.01)$; truncated at 0. and 0.1 \\
Distance & dist & pc & GDR3 & $\mathcal{U}(-5\sigma, 5\sigma)$ using parallax errors from GDR3 \\
Photometric jitter & photjitter & mag & 0.03 & $\mathcal{N}(0.02, 0.01)$; truncated at 0. and 0.1\\
\\
\multicolumn{5}{c}{\emph{TP-mode parameters}} \\
Effective temperature & Teff & K & 5000 & $\mathcal{U}(2500, 12000)$\\
Surface gravity & log(g) & dex & 4.5 & $\mathcal{U}(0.5, 5.5)$\\
Iron abundance & [Fe/H] & dex & 0.0 & $\mathcal{U}(-4, 0.5)$ \\
Alpha enrichment & [a/Fe] & dex & 0.0 & $\mathcal{U}(-0.2, 0.6)$ \\
Stellar radius (log10) & log(R) & dex; $R$ in R$_\odot$ & 0.0 & $\mathcal{U}(-3, 3)$ \\
\\
\multicolumn{5}{c}{\emph{MS-mode parameters}} \\
Equivalent evolutionary phase & EEP & --- & 250 & $\mathcal{N}(250, 50)$; truncated at 200 and 600\\
Stellar age (log10) & log(Age) & dex; Age in years & Latent variable & dsigmoid(80., 8., $-40$, 10.15, 6.0, 11.0)\\
Initial iron abundance & initial\_[Fe/H] & dex & 0.0 & $\mathcal{U}(-4, 0.5)$\\
Initial alpha enrichment & initial\_[a/Fe] & dex & 0.0 & $\mathcal{U}(-0.2, 0.6)$\\
Initial stellar mass & initial\_Mass & M$_\odot$ & 1.0 & $\mathcal{U}(0.5, 3.0)$
\enddata
\centering{
}
\end{deluxetable*}

\newpage
\movetabledown=0.5cm
\begin{longrotatetable}
\begin{deluxetable*}{lrrrrrrrr}
\label{tab:results}
\tabletypesize{\small}
\tablecolumns{7}
\tablewidth{\textwidth}
 \tablecaption{Results of performance benchmarking tests to extract fundamental parameters from TRES spectra}
 \tablehead{
 \colhead{Comparison \vspace{-0.1cm}} & 
 \colhead{$T_{\rm eff}$} &
 \colhead{$\log g$} &
 \colhead{[Fe/H]} &
 \colhead{[\textalpha/Fe]$_1$} &  
 \colhead{[\textalpha/Fe]$_2$} &
 \colhead{[\textalpha/Fe]$_3$}
 \\ 
 \colhead{} &
 \colhead{[K]} &
 \colhead{[dex]} &
 \colhead{[dex]} &
 \colhead{[dex]} &
 \colhead{[dex]} &
 \colhead{[dex]}}
\startdata
\multicolumn{7}{c}{\emph{Dwarf stars}} \\
GBS3 (S24) & 103 (128/111) & 0.090 (0.17\phantom{0}/0.058) & 0.041 (0.045/0.098) & ---& --- & --- \\
GBS3 (C25) & --- & --- & 0.054 (0.064/0.095) & 0.072 (0.072/---) & 0.11\phantom{0} (0.099/---) & 0.099 (0.087/---)\\
APOGEE & 107 (116/138) & 0.066 (0.11\phantom{0}/0.090) & 0.042 (0.041/0.091) & --- & 0.039 (0.037/---) & 0.074 (0.076/---) \\
SPOCS (VF05) & \phantom{0}99 (\phantom{0}94/\phantom{0}65) & 0.11\phantom{0} (0.15\phantom{0}/0.11\phantom{0}) & 0.061 (0.061/0.10\phantom{0}) & 0.098 (0.098/---) & 0.089 (0.088/---) & --- \\
SPOCS (B16) & \phantom{0}65 (\phantom{0}74/\phantom{0}68) & 0.055 (0.055/0.083) & 0.042 (0.042/0.12\phantom{0}) & 0.075 (0.078/---) & 0.083 (0.081/---) & 0.097 (0.10\phantom{0}/---)\\
Hyades ($\geq 0.82$M$_\odot$) & --- & --- & 0.040 (0.052/0.035) & 0.012 (0.016/---) & --- & --- \\
Asteroids & \phantom{0}17 (\phantom{0}16/\phantom{0}18) & 0.022 (0.008/0.044) & 0.044 (0.044/0.060) & 0.034 (0.036/---) & --- & --- & \\
Wide binaries ($T_{\rm eff} \geq 5000$~K) & --- & --- & 0.025 (0.034/0.049) & 0.017 (0.012/---) & --- & --- & \\
\\
\multicolumn{7}{c}{\emph{Evolved stars}} \\
GBS3 (S14) &\phantom{0}96 (\phantom{0}89/271) & 0.14\phantom{0} (0.15\phantom{0}/0.44) & 0.13\phantom{0} (0.15\phantom{0}/0.13\phantom{0}) & ---& --- & --- \\
GBS3 (C15) & --- & --- & 0.078 (0.071/0.086) & 0.22 (0.22/---) & 0.097 (0.096/---)& 0.10\phantom{0} (0.079/---) \\
APOGEE & \phantom{0}85 (\phantom{0}85/207) & 0.16\phantom{0} (0.16\phantom{0}/0.34) & 0.10\phantom{0} (0.10\phantom{0}/0.11\phantom{0}) & --- & 0.21\phantom{0} (0.20\phantom{0}/---) & 0.15\phantom{0} (0.15\phantom{0}/---) \\
SPOCS (VF05) & 158 (159/137) & 0.43\phantom{0} (0.43\phantom{0}/0.19) & 0.19\phantom{0} (0.19\phantom{0}/0.13\phantom{0}) & 0.20 (0.20/---) & 0.36\phantom{0} (0.35\phantom{0}/---) & --- \\
SPOCS (B16) & \phantom{0}36 (\phantom{0}31/264) & 0.044 (0.028/0.35) & 0.12\phantom{0} (0.11\phantom{0}/0.056) & 0.14 (0.14/---) & 0.28\phantom{0} (0.28\phantom{0}/---) & 0.19\phantom{0} (0.19\phantom{0}/---) \\
\enddata
\vspace{0.2cm} 
\tablecomments{The main entry in each cell is the uncertainty for  \texttt{uberMS} in MS mode, followed by TP-mode/\texttt{SPC} results in brackets. In most cases, the reported uncertainty is the $\sigma$MAD (1.48$\times$MAD) as this estimate is more robust to outliers than the RMS; the one exception is the asteroids, for which the reported value is simply the difference between our measurement and the nominal parameters for the Sun.  In cases where the literature source provides multiple abundance measurements for alpha elements, we list [Ti/Fe] as [\textalpha/Fe]$_1$, [Si/Fe] as [\textalpha/Fe]$_2$, and [Mg/Fe] as [\textalpha/Fe]$_3$.}
\end{deluxetable*}
\end{longrotatetable}

\onecolumngrid 
\section{Supplemental inter-method comparisons}
\label{sec:app2}

\begin{figure*}[h]
\centering
    \includegraphics[width=0.8\textwidth]{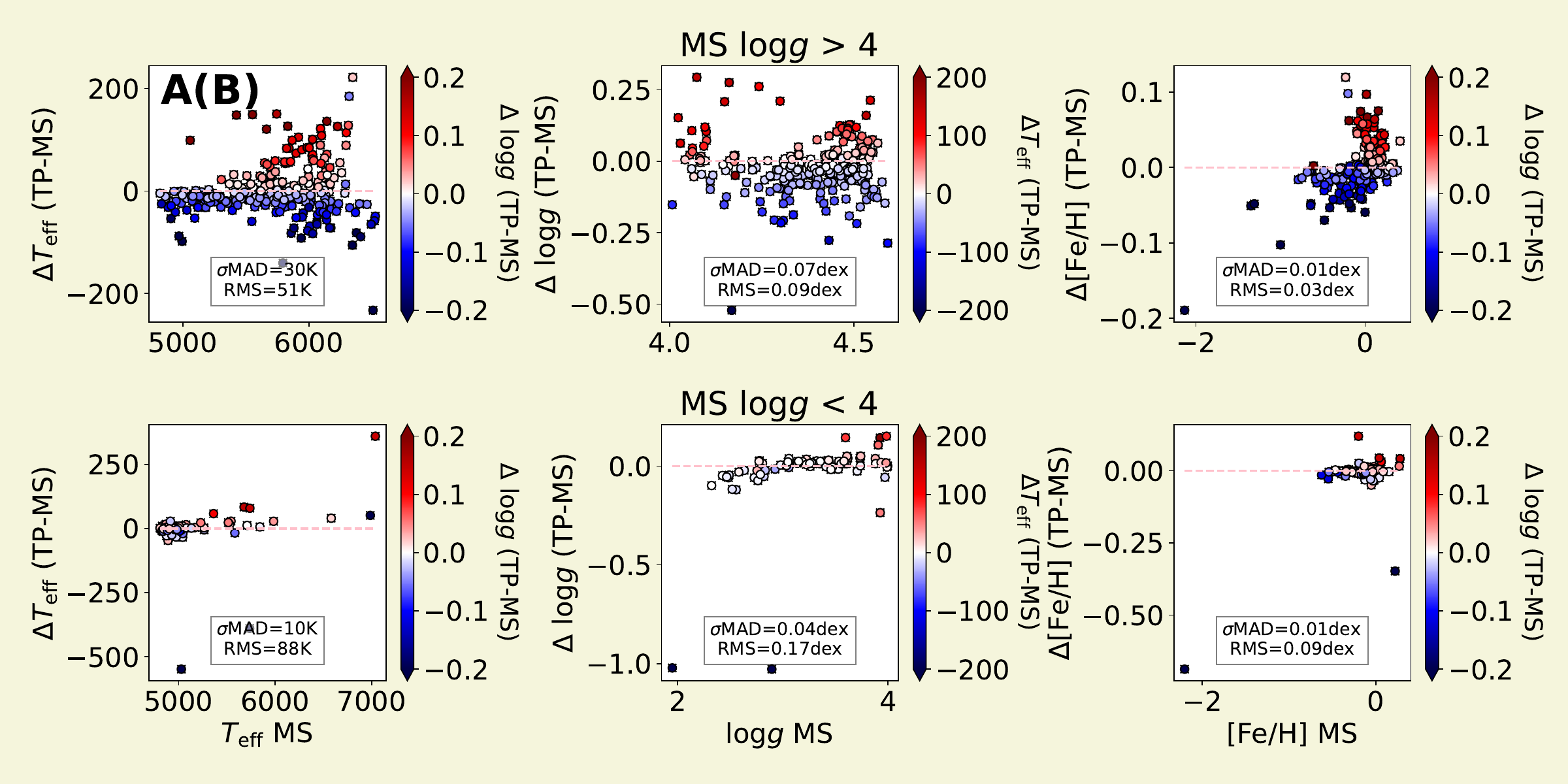}
    \includegraphics[width=0.8\textwidth]{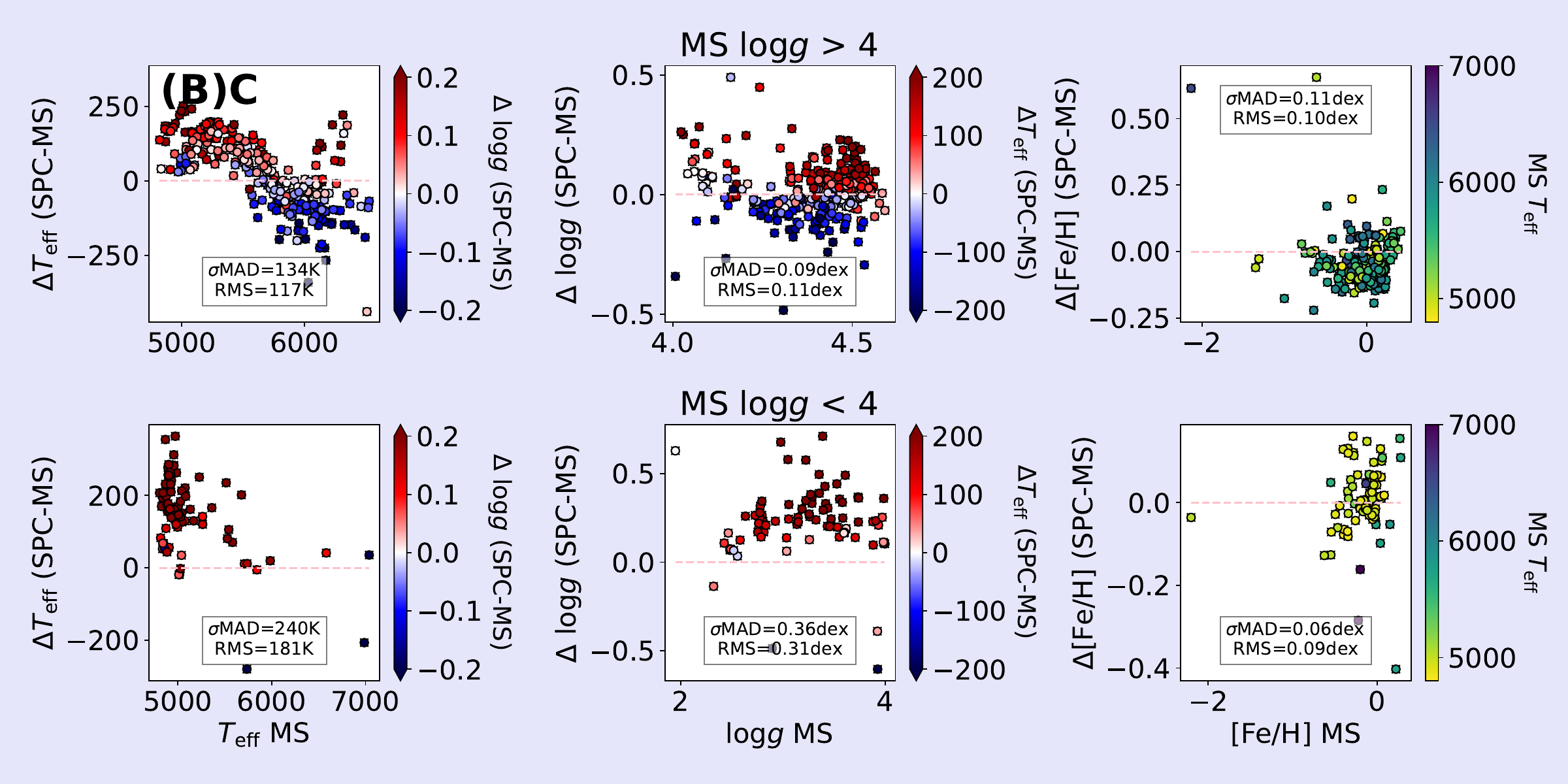}
    
    \caption{Comparison between properties estimated from TRES spectra. The upper panel A(B) compares the two \texttt{uberMS} modes, while the lower panel (B)C compares MS mode to \texttt{SPC}. In each set of subplots, the upper row shows dwarf stars (here approximated by log$g>4$ dex), while the lower row shows evolved stars (log$g<4$ dex). The dashed lines indicate unity. Note that in the upper panels, we use $\Delta\log g$ as the color bar for the [Fe/H] residual plots instead of $T_{\rm eff}$: since the main difference between the modes is the constraint on $\log g$ from the isochrone prior, we are unsurprised to find that the differences between the modes' [Fe/H] estimates are correlated with $\Delta\log g$.}
    \label{fig:supp}
\end{figure*}

\twocolumngrid
In this work, we compare the two \texttt{uberMS} modes, TP and MS, and the independent method \texttt{SPC} to a variety of benchmark samples. Comparison between the three methods is therefore implicit throughout. However, to better highlight the differences between the methods, here we combine together all the samples and provide a direct comparison between the two \texttt{uberMS} modes, and between MS mode and \texttt{SPC} (Figure~\ref{fig:supp}). We neglect stars with $T_{\rm eff} < 4800$~K from this comparison, as all three methods perform poorly in the low-temperature regime; if the cool stars are included, they dominate the plots and make it difficult to compare performance in useful regions of parameter space.

The differences between TP mode and MS mode are generally small: for dwarf stars, we find $\sigma$MADs of 30~K in $T_{\rm eff}$, 0.07~dex in $\log g$, and 0.01~dex in [Fe/H] and [\textalpha/Fe], in all cases smaller than the overall systematic uncertainties we have inferred for our MS-mode measurements of 100~K, 0.09~dex, 0.04~dex, and 0.03~dex. The most significant difference lies in $\log g$, which is unsurprising: the distinguishing feature of MS mode is that its $\log g$ measurements are informed by the isochrones, and hence they are more accurate than TP mode, which relies solely on spectrophotometry. These changes in $\log g$ propagate into modest changes in $T_{\rm eff}$, [Fe/H], and [\textalpha/Fe], as the parameters are all correlated.

The differences are larger between MS mode and \texttt{SPC}, both in terms of larger $\sigma$MADs, and also the presence of systematic offsets. For dwarf stars, \texttt{SPC} estimates hotter temperatures for cool stars and cooler temperatures for hot stars relative to \texttt{uberMS}. \texttt{SPC} also produces modestly lower metallicity estimates for dwarf stars, with a median offset of 0.067~dex; we observed similar offsets in our comparisons between \texttt{SPC} and other surveys (GBS3, APOGEE, and SPOCS). For evolved stars, the \texttt{SPC} estimates are systematically hotter and have higher surface gravities than the \texttt{uberMS} results, although the metallicities do not show a systematic offset.

\bibliography{payne}{}
\bibliographystyle{aa_url}
\end{document}